%% file: main.tex
\documentclass[sigconf]{acmart}
\usepackage[dvipsnames]{xcolor}  % 打开更多预定义颜色（包括 ForestGreen）
\definecolor{ForestGreen}{RGB}{34,139,34}
\definecolor{grey}{gray}{0.95}

\definecolor{navyblue}{HTML}{0071BC}
\definecolor{hotpink}{HTML}{FF0080}

\definecolor{oai-white}{HTML}{FFFFFF}
\definecolor{oai-black}{HTML}{000000}
\definecolor{oai-red}{HTML}{FF4500}
\definecolor{oai-green}{HTML}{51DA4C}
\definecolor{oai-blue}{HTML}{0000FF}
\definecolor{oai-yellow}{HTML}{FFF639}
\definecolor{oai-magenta}{HTML}{FF45FF}
\definecolor{oai-cyan}{HTML}{00FFFF}
\definecolor{oai-orange}{HTML}{FE7600}
\definecolor{oai-violet}{HTML}{8A2BE2}
\definecolor{oai-brown}{HTML}{A0522D}

\definecolor{oai-green-050}{HTML}{F4FFF4}
\definecolor{oai-green-100}{HTML}{E9FFE8}
\definecolor{oai-green-200}{HTML}{D9FFD8}
\definecolor{oai-green-300}{HTML}{C9FFC7}
\definecolor{oai-green-400}{HTML}{A6FFA3}
\definecolor{oai-green-500}{HTML}{7CF178}
\definecolor{oai-green-600}{HTML}{51DA4C}
\definecolor{oai-green-700}{HTML}{3FA93B}
\definecolor{oai-green-800}{HTML}{2D712A}
\definecolor{oai-green-900}{HTML}{193718}

\definecolor{oai-gray-000}{HTML}{FFFFFF}
\definecolor{oai-gray-100}{HTML}{FAFAFA}
\definecolor{oai-gray-200}{HTML}{F5F5F5}
\definecolor{oai-gray-300}{HTML}{E5E5E5}
\definecolor{oai-gray-400}{HTML}{FFB7A4}
\definecolor{oai-gray-500}{HTML}{CDCDCD}
\definecolor{oai-gray-600}{HTML}{A8A8A8}
\definecolor{oai-gray-700}{HTML}{747474}
\definecolor{oai-gray-800}{HTML}{393939}
\definecolor{oai-gray-900}{HTML}{000000}

\definecolor{visual}{HTML}{A50E0E}
\definecolor{linguistic}{HTML}{174EA6}
\definecolor{relational}{HTML}{E37400}
\definecolor{egocentric}{HTML}{0D652D}

\definecolor{tb}{HTML}{C3E4F5}
\definecolor{tg}{HTML}{D9F3CE}
\definecolor{ty}{HTML}{FFF0C5}

\definecolor{myblue}{rgb}{1.0, 0.745, 0.478}

\colorlet{mapcolor}{ForestGreen}

\newcommand{\provicon}[2]{%
  \raisebox{-0.18\height}{\includegraphics[height=2.1ex]{#1}}%
  \hspace{0.05em}#2%
}

\newcommand{\ProvOpenAI}{\provicon{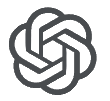}{OpenAI}}
\newcommand{\ProvGoogle}{\provicon{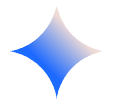}{Google}}
\newcommand{\ProvAnthropic}{\provicon{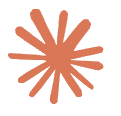}{Anthropic}}
\newcommand{\ProvDeepSeek}{\provicon{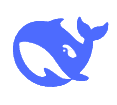}{DeepSeek}}

\usepackage{amsmath,amssymb}
\usepackage{bbm}
\usepackage{textcomp}
\DeclareUnicodeCharacter{2212}{-} % 讓 Unicode 負號變成普通負號

\usepackage{graphicx}
\usepackage{adjustbox}
\usepackage{wrapfig}

\usepackage{booktabs}
\usepackage{tabularx}
\usepackage{array}
\usepackage{multirow}
\usepackage{longtable}
\usepackage{makecell}
\usepackage{colortbl}

\usepackage{caption}
\usepackage{subcaption}

\usepackage{algorithm}
\usepackage{algorithmicx}
\usepackage{algpseudocode}

\usepackage{balance}
\usepackage{enumitem}

\usepackage{pgf}
\usepackage{tikz}
\usepackage{forest}
\usepackage{xspace}

\usepackage[most]{tcolorbox}
\usepackage{csquotes}
\usepackage[noorphans,vskip=1em,leftmargin=1em]{quoting}

\usepackage{xurl}

\hypersetup{
  colorlinks=true,
  linkcolor=black,
  citecolor=black,
  urlcolor=oai-red,
}

\usepackage{pifont}     % \ding 用来画 √ ×

\newcommand{\bench}{\texttt{RAL-Bench}\xspace}
\newcolumntype{Y}{>{\raggedright\arraybackslash}X}
\newcolumntype{R}[1]{>{\raggedleft\arraybackslash}p{#1}}
\setcopyright{none}

\usepackage{siunitx}
\definecolor{myitem}{HTML}{558BBB}

\newcommand{\cnum}[2]{%
  \tikz[baseline=(C.base)]%
    \node[draw=none, circle, fill=#1, inner sep=1.3pt, minimum size=1em] (C)
    {\color{white}\bfseries\scriptsize #2};%
}

\newcommand{\circnum}[1]{%
  \tikz[baseline=(n.base)]{
    \node (n) [circle, fill=black, text=white, font=\bfseries\scriptsize, inner sep=1.2pt] {#1};
  }%
}

\begin{document}

% \pagenumbering{arabic}
% \thispagestyle{plain}
% \pagestyle{plain}

% \title{Thinking Beyond Snippets: Are Large Language Models Ready for Real Application-Level Code Generation? }

% \title{RAL-Bench: An Executable System-Level Benchmark for Requirement Satisfaction and Software Quality}

% \title{Thinking Beyond Runnable: Are LLMs Ready for Application-Level Functional Correctness and Non-functional Quality?}

% \title{RAL-Bench: Benchmarking Application-Level Code Generation with Functional and Non-Functional Evaluation}

\title{Toward Functional and Non-Functional Evaluation of Application-Level Code Generation}

% \author{
% Ruwei Pan\textsuperscript{1} \quad
% Yakun Zhang\textsuperscript{2} \quad
% Lu Zhang\textsuperscript{3} \quad
% Hongyu Zhang\textsuperscript{1,$\dagger$} \\
% \textsuperscript{1}Chongqing University \quad
% \textsuperscript{2}Harbin Institute of Technology \quad
% \textsuperscript{3}Peking University \\
% \textsuperscript{$\dagger$}Corresponding author.
% }

\newcommand{\inst}[1]{\textsuperscript{#1}}

\settopmatter{authorsperrow=1}

\author{
Ruwei Pan\inst{1,2} \quad
Yakun Zhang\inst{3} \quad
Qingyuan Liang\inst{2} \quad
Yueheng Zhu\inst{1,2} \quad
Chao Liu\inst{1} \quad
Lu Zhang\inst{2} \quad
Hongyu Zhang\inst{1}
}

\authornote{
\inst{1} Chongqing University \quad
\inst{2} Peking University \quad
\inst{3} Harbin Institute of Technology (Shenzhen)
}

%\settopmatter{printacmref=false} 

% \begin{CCSXML}
% <ccs2012>
%  <concept>
%   <concept_id>10010520.10010553.10010562</concept_id>
%   <concept_desc>Computer systems organization~Embedded systems</concept_desc>
%   <concept_significance>500</concept_significance>
%  </concept>
%  <concept>
%   <concept_id>10010520.10010575.10010755</concept_id>
%   <concept_desc>Computer systems organization~Redundancy</concept_desc>
%   <concept_significance>300</concept_significance>
%  </concept>
%  <concept>
%   <concept_id>10010520.10010553.10010554</concept_id>
%   <concept_desc>Computer systems organization~Robotics</concept_desc>
%   <concept_significance>100</concept_significance>
%  </concept>
%  <concept>
%   <concept_id>10003033.10003083.10003095</concept_id>
%   <concept_desc>Networks~Network reliability</concept_desc>
%   <concept_significance>100</concept_significance>
%  </concept>
% </ccs2012>
% \end{CCSXML}

% \ccsdesc[500]{Software and its engineering~Software development techniques}


\input{Abstract}

\maketitle

\input{Introduction}

\input{Approach}

\input{Evaluation}

\input{Discussion}

\input{Related_Work}

\input{Conclusion_and_Future_Work}

\balance
\bibliographystyle{ACM-Reference-Format}
\bibliography{references}
\balance

\end{document}

%% file: Abstract.tex
\begin{abstract}

Large language models (LLMs) have achieved strong performance on code generation. However, most prior evaluations focus on snippet-level outputs, such as function generation or repository completion. These settings do not fully evaluate application-level code generation, where the goal is to produce a runnable repository with coherent multi-file structure, dependency support, and end-to-end executability. In addition, real-world software quality depends not only on functional correctness but also on non-functional quality attributes, such as maintainability and security. In this paper, we present \bench, a benchmark and evaluation framework for application-level code generation. For each task, \bench derives a concise natural-language requirement from a high-quality reference project, constructs black-box system tests for both functional correctness and non-functional quality attributes. It also retains only the candidate tests that pass on the reference repository. 
Under this unified evaluation protocol, functional correctness is measured by the system test pass rate, while non-functional quality is evaluated along five ISO/IEC 25010-inspired dimensions, with per-dimension diagnostics and reference-normalized scoring.
We evaluate 16 frontier LLMs under a controlled zero-shot setting with greedy decoding. The results show that functional correctness remains the primary bottleneck in application-level code generation, while non-functional quality also remains challenging. 
Under our evaluation protocol, no model exceeds a 45\% functional score. These findings suggest that strong performance on existing code generation benchmarks does not yet translate to strong performance on application-level repository generation. This result highlights the need for evaluation settings that directly assess end-to-end repository generation rather than relying only on snippet-level success.
% To support future research, \bench is available at \href{https://anonymous.4open.science/r/RAL-Bench}
%      {\textbf{\textcolor{oai-blue}{our anonymous repository}}}.
% We release \bench\ at
% \href{https://anonymous.4open.science/r/RAL-Bench}
%      {\textbf{\textcolor{oai-blue}{our anonymous repository}}}
% to support future research on LLM-based code generation.

\end{abstract}

% \hy{I feel that the paper title  does not reflect the paper well. "Thinking Beyond Snippets" - research on repo-level code generation level has been there for many years. Code generation at repo/applicatio level is nothing new. Many current AI coding tools can do it too}
% \rw{Is the current title ok?}

%% file: Introduction.tex
\section{Introduction}

% In daily development, we often articulate our needs in the form of a single, highly compressed requirement like \textit{"implement a pure Python steganography library"}. 
% For human developers, the real difficulty usually does not lie in implementing a particular function, but in integrating it into a fully working application \citep{mukherjee2021fixing, rausch2017empirical}. 
% This process requires decisions about project organization, dependency management, functional completeness, and compliance with non-functional quality attributes such as robustness requirements \citep{kazman2015case, cataldo2009software, eckhardt2016non}.
% % In contrast, code produced by LLMs frequently demonstrates a different characteristic. This process presents code as a seemingly complete project but fails during dependency installation or runtime usage, 
% % failure modes that snippet-level evaluations typically do not capture.
% In contrast, LLM-generated code can appear as a complete project while failing during dependency installation or runtime execution, which snippet-level evaluations typically do not capture.

In practice, software needs are often expressed as short natural-language requests, such as \textit{"implement a pure Python steganography library"}. Turning such a request into a runnable application requires not only implementing functions, but also making decisions about project structure, dependencies, interfaces, and runtime behavior \citep{mukherjee2021fixing, rausch2017empirical, kazman2015case, cataldo2009software, eckhardt2016non}. A repository may therefore look complete but still fail during installation or execution. Such failures are usually not exposed by snippet-level evaluation.

\begin{figure}[t]
  \centering
   \includegraphics[width=0.45\textwidth]{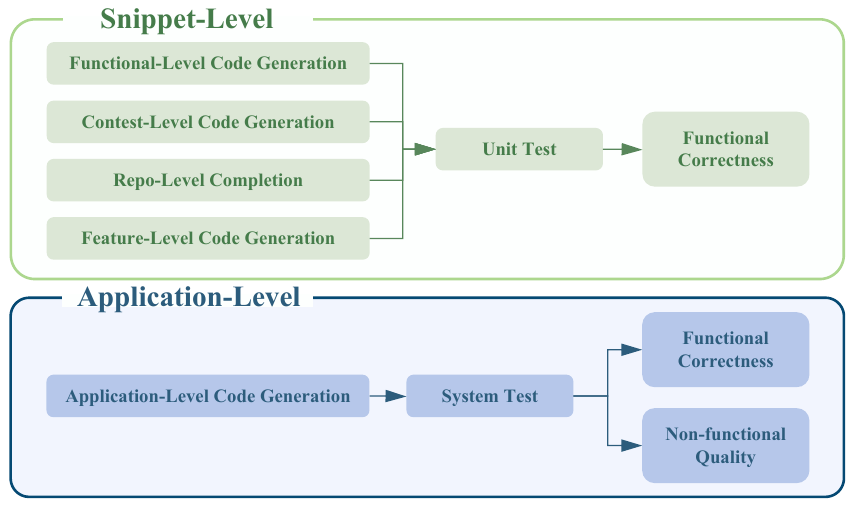}
  \caption{Scope comparison between snippet-level and application-level code generation.}
  \label{fig:levels}
  \vspace{-7px}
\end{figure}

Recent code LLMs have made strong progress on snippet-level tasks, and much of this progress has been established on widely used code generation benchmarks \citep{chen2022codet, li2023starcoder, roziere2023code}. These benchmarks mainly study function-level, contest-level, repo-level, or feature-level generation, and most of them evaluate functional correctness with unit tests \citep{chen2021evaluating, austin2021program, hendrycks2021measuring, jain2024livecodebench, yu2024codereval, li2024deveval, deng2025nocode}. While useful for measuring snippet-level ability, they do not fully evaluate application-level generation. At the application level, an LLM must produce a runnable repository that can be installed, executed, and checked end to end. This setting also requires evaluation beyond functional correctness, because a usable application must satisfy basic non-functional quality attributes. Figure~\ref{fig:levels} illustrates this difference. This gap leads to two limitations in current benchmarks.

\begin{itemize}
    % \item \textbf{Lack of application-level evaluation.} Most benchmarks focus on evaluating functional correctness using snippet-level outputs under predefined unit tests. They do not test whether an LLM can generate a runnable project with proper structures, dependency management and end-to-end executability.
    \item \textbf{Lack of application-level evaluation.} Most benchmarks do not test whether an LLM can generate an executable application from a natural-language requirement.

    % \item \textbf{No assessment of non-functional quality attributes.} Applications are constrained not only by functional correctness but also by 
    % % maintainability, security, robustness, efficiency and resource usage
    % non-functional quality attributes. However, existing benchmarks almost exclusively focus on functional tests. 
    % While a few snippet-level benchmarks have started to consider non-functional aspects, they focus on a single attribute (e.g., efficiency or security) on isolated code snippets, rather than assessing multi-attribute non-functional quality end to end.
    % % Consequently, they often miss non-functional failures such as excessive latency or resource mismanagement.
    % Real-world application quality is a multi-attribute construct that cannot be reduced to functional correctness alone. 
    % An implementation may pass functional tests yet remain unusable in practice due to latency, resource inefficiency, robustness issues, security risks or poor maintainability.

    \item \textbf{Limited end-to-end non-functional evaluation.} Most benchmarks do not assess non-functional quality for complete applications. Some recent work considers a single attribute, such as efficiency or security, on isolated snippets. This is still different from evaluating multi-attribute quality for end-to-end applications.

% , thus failing to capture runtime failures such as excessive latency or resource mismanagement.
\end{itemize}

These limitations leave an important evaluation gap. Current benchmark results do not tell us how well LLMs support application-level code generation. This paper studies a focused question: \textit{Can current LLMs generate application-level repositories that satisfy both functional correctness and non-functional quality attributes?}

% \textbf{Our proposal.} 
% % In this work, we set out to answer the important question and \textit{evaluate} the evaluation dataset. 
% In this work, we set out to answer this question by revisiting how evaluation datasets are constructed and by building a more realistic evaluation framework.
\textbf{Benchmark overview.} Our goal is to introduce a benchmark for studying application-level code generation.
% Consequently, we build \bench - an evaluation framework to assess LLMs’ ability to generate real, end-to-end executable applications from natural language requirement. 
Specifically, we introduce \bench (\textbf{R}eal-World \textbf{A}pplication-\textbf{L}evel Code Generation \textbf{Bench}), a benchmark and evaluation framework for application-level code generation. For each task, we derive a concise requirement from a real GitHub project, build black-box system tests for functional and non-functional evaluation. We then validate the candidate tests on the reference repository before applying them to generated code. This design keeps the benchmark close to real software usage while maintaining a controlled evaluation protocol.

\textbf{Contributions.} This paper makes three contributions.

\begin{itemize}
    % \item \textbf{Study:} We are the first to systematically analyze the evaluation gap between existing code benchmarks and the requirements of real-world application development. Our study also opens up a new research direction for precisely and rigorously evaluating application-level code generation.
    \item We define application-level code generation as a benchmark setting that requires executable repositories under end-to-end evaluation, rather than snippet-level outputs under unit tests.
    
    % \item \textbf{Approach:} We propose \bench, a benchmark and evaluation framework for application-level code generation grounded in real-world GitHub repositories. For each task, we extract a concise natural-language requirement from a high-quality reference project and construct black-box system tests covering both functional correctness and key non-functional quality attributes. We execute all candidate tests on the reference repository and retain only those that pass, ensuring a sound test oracle and end-to-end executability. Functional score is computed as the system test pass rate. Non-functional quality is measured along five ISO/IEC 25010-inspired dimensions and aggregated using an AHP-derived weight vector with per-dimension diagnostics \citep{botchway2021evaluating}. In addition, baseline non-functional metrics are collected on the reference repository to enable baseline-normalized scoring.

    \item We present \bench, which uses real GitHub projects, reference-validated black-box system tests, and multiple dimensional non-functional evaluation for application-level code generation.

    \item We provide a controlled empirical study of 16 frontier LLMs under a zero-shot setting. The results show that functional correctness remains the main bottleneck in application-level code generation.
    
    % \item \textbf{Results:} We comprehensively evaluate 16 LLMs (standard and thinking) under zero-shot settings with greedy decoding. 
    % First, we find that functional correctness is the dominant bottleneck: under our requirement-driven, reference-validated black-box system tests, no LLM surpasses a 45\% functional pass rate.
    % Second, although non-functional scores are generally higher, they cannot offset functional failures.
    % Third, our failure-pattern dataset comprises 446 successfully generated repositories and over 4,500 test-case execution logs. It shows that failures are dominated by Requirement–Implementation Mismatch and Non-functional Quality Failures (82.8\% combined), whereas Executability \& Dependency Failures account for 17.2\%. 
    % Fourth, we quantify cost. Thinking LLMs are more expensive on average, yet they do not yield consistent functional improvements. This suggests that higher-cost “thinking” does not yet translate into effective reasoning for application-level generation.
    % Finally, the results show that when tasks scale to the application level, mainstream code generation strategies are no longer effective.
    
\end{itemize}

%% file: Approach.tex
\section{ \bench }

\subsection{Overview}
% We introduce \bench  to quantitatively evaluate the ability of LLMs to perform application-level code generation. 
% \bench comprises over 450 functional and non-functional evaluation points derived from 38 real projects and covers seven distinct real-world usage scenarios (e.g., Data and Security), as summarized in Table \ref{tab:repo-manifest-scenarios}.
We introduce \bench as a benchmark and evaluation protocol for application-level code generation. Our goal is controlled comparison of LLMs under a unified generation and testing setting, rather than best-case performance under task-specific prompting or agent design. \bench contains 38 real projects, more than 450 evaluation points, and seven application scenarios, as summarized in Table \ref{tab:repo-manifest-scenarios}.
These projects are sourced from actively maintained GitHub repositories with an average popularity of more than 1,000 stars, ensuring that they have documented functionality and real usage.
% Each project provides clearly specified functional requirements and measurable non-functional quality attributes (e.g., efficiency and security). 
% Moreover, none of the projects are drawn from existing code generation benchmarks or public evaluation suites, which helps mitigate potential benchmark contamination and data leakage for LLMs.
We exclude repositories that are already used in public code generation benchmarks or evaluation suites, which reduces benchmark contamination. However, because the repositories are public, we do not claim that pretraining exposure can be fully ruled out. We therefore interpret the benchmark as measuring application-level generation under realistic public-repository exposure conditions.
% \bench is iteratively reviewed to minimize requirement ambiguity and to ensure that all manually designed test cases are precise, suitable, and executable for rigorous evaluation.
For each task, the requirement and test suite are manually reviewed to reduce ambiguity. We remove project-specific wording that reveals implementation details, and verify that all retained tests execute successfully on the pinned reference repository.

% \bench includes ... tasks of  ... types: ... See Fig. \ref{} for an overview of \bench tasks and Fig. \ref{} for dataset statistics.

% \begin{figure}[t]
%   \centering
%   \includegraphics[width=0.5\columnwidth]{Fig/scenario_pie_counts_thick.pdf}
%   \caption{Scenario distribution across seven application domains (n=38).\hy{perhaps a table is better here, or just use texts}}
%   \label{fig:scenario-dist}
% \end{figure}

% Requires: \usepackage{booktabs}
% \begin{table}[t]
%   \centering
%   \caption{Scenario distribution.}
%   \label{tab:scenario-share-2rows}
%   \setlength{\tabcolsep}{5pt}
%   \renewcommand{\arraystretch}{1.08}
%   \small
%   \begin{tabular}{lccccccc}
%     \toprule
%     \textbf{Scenario} & \textbf{Tooling} & \textbf{Data} & \textbf{Web} & \textbf{Security} &
%     \textbf{Automation} & \textbf{Observability} & \textbf{Content} \\
%     \midrule
%     Share (\%)         & 15.8 & 15.8 & 10.5 & 13.2 & 15.8 & 13.2 & 15.8 \\
%     \bottomrule
%   \end{tabular}
% \end{table}

% Requires: \usepackage{booktabs,multirow}
% in preamble:
% \usepackage{booktabs}
% \usepackage{multirow}
% \usepackage{graphicx}

\begin{table}[t]
\centering
\small
\setlength{\tabcolsep}{6pt}
\renewcommand{\arraystretch}{1.08}
\caption{Repository manifest grouped by the seven real-world usage scenarios. Pinned commit SHA fixes the exact repository snapshot for reproducible evaluation.}
\label{tab:repo-manifest-scenarios}
\resizebox{\columnwidth}{!}{%
\begin{tabular}{l l p{6.0cm} r r}
\toprule
\textbf{Scenario} & \textbf{Repository} & \textbf{Commit (40-hex SHA)} & \textbf{\#Files} & \textbf{LoC} \\
\midrule
\multirow{6}{*}{Tooling} & astanin/python-tabulate & {\ttfamily 74885be915e2ac611585f5398f23c402532c1059} & 1 & 2.4k \\ % intent: tabular output formatting for CLI/reporting
 & martinblech/xmltodict & {\ttfamily f7d76c96fc0141238947abcc5fa925d3ffd9eb78} & 3 & 1.4k \\ % intent: XML <-> dict/JSON conversion utility
 & mkaz/termgraph & {\ttfamily b86ccfddf55eda83bb2c3b9afe4c74478559bcb1} & 8 & 1.0k \\ % intent: terminal visualization tool
 & pallets/click & {\ttfamily cdab890e57a30a9f437b88ce9652f7bfce980c1f} & 16 & 7.7k \\ % intent: building command-line interfaces
 & pygments/pygments & {\ttfamily 28ec10c5e154ee27201997feb573a0b065f74082} & 339 & 109.5k \\ % intent: syntax highlighting/formatting infrastructure
 & python-cmd2/cmd2 & {\ttfamily 926636a38156e4a7e07dc1cee04bca7b019776ec} & 21 & 9.1k \\ % intent: interactive CLI shells
\midrule
\multirow{6}{*}{Data} & CamDavidsonPilon/lifelines & {\ttfamily 47afb1c1a272b0f03e0c8ca00e63df27eb2a0560} & 47 & 20.8k \\ % intent: survival analysis/statistical modeling on datasets
 & dateutil/dateutil & {\ttfamily e081f6725fbb49cae6eedab7010f517e8490859b} & 37 & 14.4k \\ % intent: datetime parsing/computation as data primitive
 & jazzband/tablib & {\ttfamily 7d6c58a574782b2525af17225e93bdf4efa0f376} & 31 & 5.0k \\ % intent: dataset import/export across formats
 & petl-developers/petl & {\ttfamily 482fc04fd2589ac9404a8e6a21601b956aa64a2f} & 124 & 27.0k \\ % intent: ETL pipelines for tabular data
 & pudo/dataset & {\ttfamily b2ab09e58c6f17334e4286009b20887b6a8a8fac} & 6 & 1.1k \\ % intent: database-as-dataset abstraction
 & python-pendulum/pendulum & {\ttfamily 754ed58a6c232b2335a967d5e266fc48c448b8f3} & 172 & 19.6k \\ % intent: advanced datetime arithmetic
\midrule
\multirow{4}{*}{Web} & Python-Markdown/markdown & {\ttfamily e5fa5b86e8ec380cbc520cfc637d72c779e5c601} & 33 & 5.7k \\ % intent: render Markdown to HTML for web publishing
 & fastapi/sqlmodel & {\ttfamily 1e4bf5c190e849e5d070f437a061667776788013} & 12 & 2.3k \\ % intent: web-backend modeling (FastAPI ecosystem)
 & psf/requests & {\ttfamily 70298332899f25826e35e42f8d83425124f755a5} & 35 & 8.4k \\ % intent: HTTP client for web services/APIs
 & python-visualization/folium & {\ttfamily 0ff5d993bf2c60dbf8f56c12bbc67f104b97687c} & 48 & 8.2k \\ % intent: generate browser-facing interactive maps (HTML)
\midrule
\multirow{5}{*}{Security} & cedricbonhomme/Stegano & {\ttfamily 3ffac4782b8e4e5893c3919e1eabb921867220cb} & 18 & 1.1k \\ % intent: steganography / information hiding
 & fail2ban/fail2ban & {\ttfamily 9887ee441215a2a1f889b0e1baaa43cd3a956d26} & 78 & 24.9k \\ % intent: intrusion prevention / banning attackers
 & jpadilla/pyjwt & {\ttfamily 04947d75dc45ba1a4a66eaa2b24fbb0eb512ceab} & 12 & 2.1k \\ % intent: JWT signing/verification for auth
 & mitmproxy/mitmproxy & {\ttfamily a7621021191011de7c95771459688de3ecd67c10} & 242 & 39.1k \\ % intent: intercepting proxy for security/traffic analysis
 & sqlmapproject/sqlmap & {\ttfamily 876f14199ecb33ee16ed09b40f9ebe5225a1cd74} & 102 & 27.6k \\ % intent: SQL injection testing/exploitation
\midrule
\multirow{6}{*}{Automation} & celery/celery & {\ttfamily 4d068b5667b75e21844f0748d13627e70d5a42ac} & 156 & 32.2k \\ % intent: distributed task queue / async jobs
 & dbader/schedule & {\ttfamily 82a43db1b938d8fdf60103bd41f329e06c8d3651} & 1 & 0.7k \\ % intent: lightweight scheduling
 & fastapi/typer & {\ttfamily a8c425b613aeb645553a418b4aa6e3b7f4a8db77} & 16 & 4.1k \\ % intent: CLI-driven scripting/automation entrypoints
 & msiemens/tinydb & {\ttfamily 2283a2b556d5ef361b61bae8ed89c0ce0730d7c8} & 10 & 1.4k \\ % intent: embedded persistence for scripts
 & nvbn/thefuck & {\ttfamily c7e7e1d884d3bb241ea6448f72a989434c2a35ec} & 206 & 4.8k \\ % intent: automate correction of shell commands
 & tkem/cachetools & {\ttfamily 9983ef8bd76758707ab9d197d4bd9fa47b4fb8bd} & 5 & 1.0k \\ % intent: caching utilities used in automated workflows/services
\midrule
\multirow{5}{*}{Observability} & Delgan/loguru & {\ttfamily 764cd30d4b285ca01718ab3162a9f1a022bc49c6} & 19 & 4.0k \\ % intent: logging for observability
 & Textualize/rich & {\ttfamily 36fe3f7ca9becca4777861d5e6e625f5a4a37545} & 78 & 23.2k \\ % intent: terminal dashboards/log presentation
 & gorakhargosh/watchdog & {\ttfamily 3f8a12f9cf6d9f8091e05b61952409ed9623257d} & 54 & 8.2k \\ % intent: watch system events for monitoring/triggering
 & nicolargo/glances & {\ttfamily 7cd3eeb1a67c8a1623887e0d8b7de70797a9696e} & 129 & 18.7k \\ % intent: system monitoring
 & python-humanize/humanize & {\ttfamily c2b096a96fa8850399a84401267d5c174e57e7e3} & 12 & 2.6k \\ % intent: human-readable telemetry/value formatting
\midrule
\multirow{6}{*}{Content} & imageio/imageio & {\ttfamily 761929cc6ae6503f93f2431e0b47d504837ba77a} & 49 & 26.4k \\ % intent: image/video IO and processing
 & mailpile/Mailpile & {\ttfamily 741e610f3e765f03bd0c452c92c6758280e7f99f} & 130 & 43.2k \\ % intent: email content indexing/management
 & py-pdf/pypdf & {\ttfamily 19735763b856cccf0f69630d0f582a448ec5d8bb} & 56 & 37.3k \\ % intent: PDF document processing
 & quodlibet/mutagen & {\ttfamily 905e8152d57efc1eb80ae3cb2b557735ede8d070} & 56 & 14.5k \\ % intent: audio metadata/content management
 & sffjunkie/astral & {\ttfamily ac23ab5c0c69837d8fa1a5bb184c2d6d125b26b3} & 36 & 4.9k \\ % intent: astronomical time info for user-facing content/apps
 & un33k/python-slugify & {\ttfamily 7b6d5d96c1995e6dccb39a19a13ba78d7d0a3ee4} & 5 & 0.3k \\ % intent: text normalization/slug generation for publishing
\bottomrule
\end{tabular}%
}
\vspace{-15px}
\end{table}

\begin{figure*}[t] 
  \centering
  \includegraphics[width=0.75\textwidth]{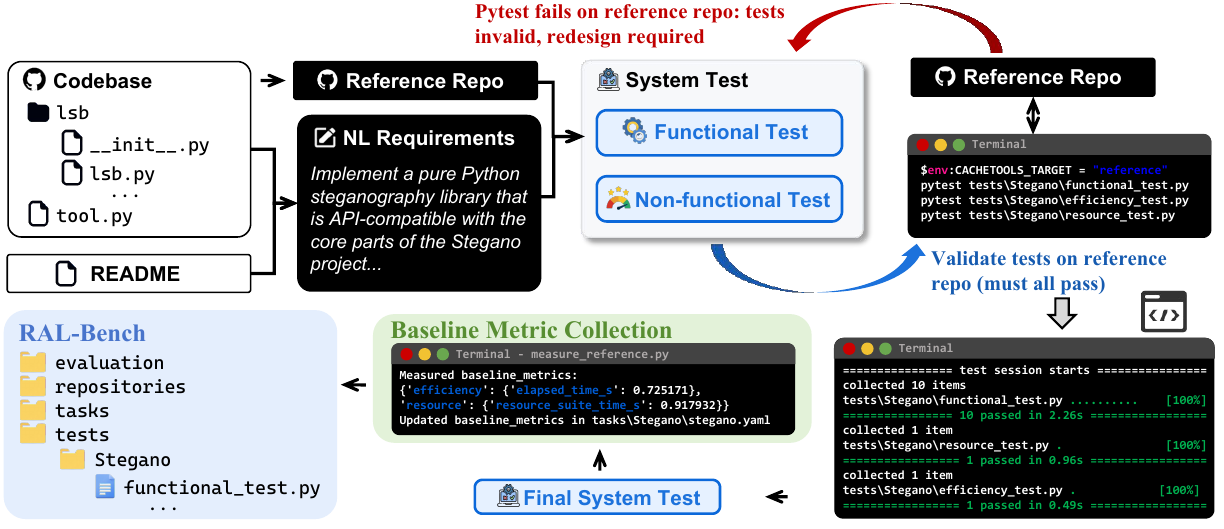}
  
  \caption{Overall construction pipeline of \bench.}
  \label{fig:construction-pipeline}
  \vspace{-7px}
\end{figure*}

\subsection{Benchmark Construction}

% We develop an executable benchmark construction pipeline to curate real-world tasks, derive natural-language requirements, build executable system test suites, and evaluate generated repositories at scale, as shown in Fig. \ref{fig:construction-pipeline}.
We build \bench through four steps: \textbf{repository selection, requirement drafting, system test construction, and reference baseline collection}. The goal of this pipeline is to produce tasks that are executable, comparable, and close to real software usage, while keeping the evaluation process reproducible.
The whole workflow is shown in Fig. \ref{fig:construction-pipeline}.

\textbf{Reference Repository Selection.} 
% We begin by selecting well-established GitHub projects as reference repositories, each with clearly documented functionality and active real-world use. For every chosen project, we pin a specific commit and snapshot its codebase within our evaluation framework. 
% We treat this pinned snapshot as the ground-truth implementation because it is a stable, widely adopted application-level project. 
We select reference repositories using three criteria: \textbf{clear documented functionality, stable public interfaces that support end-to-end testing, and active real-world usage}. We exclude repositories that are already used in public code generation benchmarks. For each selected repository, we pin one commit and use that snapshot throughout benchmark construction and evaluation.

\textbf{Natural-Language Requirement Distillation.} 
% Given a reference repository, we distill a natural-language requirement description (NL requirement) from its README and source code. First, we synthesize a concise requirement description that summarizes the project’s core intent. Second, we deliberately phrase the requirement to mirror how developers articulate intent in practice. It takes the form of a single, highly compressed request, such as \textit{implement a pure Python steganography library}. Third, the requirement specifies what the application should do and how it is expected to be used. However, it intentionally avoids prescribing a concrete project layout or implementation strategy. Under this setting, LLMs must expand a terse intent into a fully structured, executable project and infer the architecture, dependencies, and non-functional considerations without further guidance.
Given a reference repository, we draft one concise natural-language requirement from its README, public API surface, usage examples, and source code. README and examples provide the high-level task description, while the source code is used only to verify scope and expected usage patterns. We then manually revise the draft to remove implementation details, file-layout hints, and project-specific names that would reveal the answer. The final requirement states what the application should do and how it is expected to be used, but does not prescribe the internal implementation. 
The requirement itself remains compressed. During evaluation, we additionally provide the expected module and package surface of the reference repository. This interface specification is not intended to reveal implementation logic. Its role is to reduce evaluation artifacts caused by mismatched entry points or module paths.

\textbf{System Test Generation.} 
% Next, we construct system test suites for each task, consisting of black-box functional tests and non-functional tests (e.g., efficiency and security). Functional tests aim to exercise diverse functional requirements derived from real-world projects. Non-functional tests assess key non-functional quality attributes via static analysis and controlled runtime checks. Each candidate test case is first executed against the corresponding reference repository. If a test fails on the reference implementation, we deem it invalid and discard it. We retain only test cases that pass on the reference repository. This filtering ensures a sound test oracle and guarantees that the suite can be executed end-to-end in a fully automated pipeline.
We construct black-box system tests from documented behaviors, public APIs, usage examples, and expected input-output patterns of the reference project. Functional tests cover core behaviors, common usage paths, and representative error cases. Non-functional tests assess maintainability, security, robustness, efficiency, and resource usage through static analysis or controlled runtime measurement. Each candidate test is first executed on the pinned reference repository. Tests that fail, depend on unstable external conditions, or rely on undocumented behavior are removed. We retain only tests that execute successfully and produce consistent expected outcomes on the reference snapshot. This step supports oracle validity, but does not by itself guarantee full behavioral coverage. We therefore also report test statistics and coverage-oriented evidence in Section 2.5.

\textbf{Baseline Metric Collection.} 
% We run the non-functional test suite on the reference repository to collect baseline metrics for each task. These statistics are stored alongside the task configuration and later used when evaluating generated repositories. This enables fair and consistent comparisons across models and tasks. 
We run the same non-functional measurement pipeline on the reference repository and store the resulting per-task baselines. These baselines are used as task-specific anchors for cross-project comparison. They are not treated as globally optimal targets.

\subsection{Metric Design} 

We evaluate application-level code generation using two complementary metrics: functional correctness and non-functional quality attributes.

\underline{Functional correctness.} 
% The functional score of functional correctness is defined as the functional test pass rate, computed as the number of passed tests divided by the total number of tests for each task. 
Functional correctness is measured by the pass rate of the functional system tests for each task. This score captures whether the generated repository satisfies the tested end-to-end behaviors of the target application.
% Passing functional tests alone does not fully reflect the real-world usability of generated applications.

\underline{Non-functional quality attributes.} 
% Following the ISO/IEC 25010 quality model \citep{estdale2018applying}, we additionally assess five non-functional dimensions that are critical in practice: maintainability, security, robustness, efficiency, and resource usage. 
Following ISO/IEC 25010 \citep{estdale2018applying}, we evaluate five non-functional dimensions: maintainability, security, robustness, efficiency, and resource usage. We use these dimensions as scalable operational proxies for practical software quality, rather than exhaustive measurements of all quality concerns.

\noindent \cnum{myitem}{1} \textbf{Maintainability} is measured by the lower-bound Maintainability Index (MI) from static analysis. MI does not capture all aspects of maintainability, such as naming quality or architectural clarity, but it provides a low-cost and reproducible signal of structural complexity across projects. Higher MI generally reflects lower structural complexity and therefore improved maintainability. Let $g$ and $b$ denote the MI lower bounds of the generated code and the reference implementation, respectively. To avoid hard saturation when directly using $g/b$, we apply a smooth compression function:
\begin{equation}
\mathrm{M} \;=\; \frac{g/b}{1 + g/b}.
\label{eq:maintainability}
\end{equation}

\noindent \cnum{myitem}{2} \textbf{Security} is quantified by the number of high-risk issues reported by static security analysis. We use only high-risk issues to reduce noise from low-severity issues. This metric should be read as a conservative risk signal rather than a complete measure of software security, because static analysis may still produce false positives or miss domain-specific issues. Let $g$ and $b$ denote the number of high-risk findings in the generated code and the reference implementation, respectively. We compute a reference-aligned inverse ratio with smoothing:
\begin{equation}
\mathrm{S} \;=\; \min\!\left(1,\; \frac{b+1}{g+1}\right).
\label{eq:security}
\end{equation}

\noindent \cnum{myitem}{3} \textbf{Robustness} is evaluated using a dedicated robustness test suite targeting invalid, boundary, and unexpected inputs. This choice reflects runtime stability under black-box execution rather than static robustness properties. Let $\textit{passed}$ and $\textit{total}$ denote the number of passed tests and the total number of robustness tests. The robustness score is defined as:
\begin{equation}
\mathrm{Rb} \;=\; \frac{\textit{passed}}{\textit{total}}.
\label{eq:robustness}
\end{equation}

\noindent \cnum{myitem}{4} \textbf{Efficiency} is measured by the elapsed runtime of a dedicated efficiency suite under a fixed environment. Our goal is not fine-grained performance profiling, but a reproducible end-to-end efficiency signal that can be compared across generated and reference repositories. Let $T_{\text{gen}}$ and $T_{\text{ref}}$ denote the efficiency-suite execution times of the generated code and the reference implementation, respectively. If the efficiency suite fails to execute successfully (e.g., failed tests), we set $\mathrm{E}=0$; otherwise:
\begin{equation}
\mathrm{E} \;=\; \min\!\left(1,\; \frac{T_{\text{ref}}}{T_{\text{gen}}}\right).
\label{eq:efficiency}
\end{equation}

\noindent \cnum{myitem}{5} \textbf{Resource usage} is measured using a \emph{dedicated resource suite}. During execution, we sample resident set size (RSS) memory usage and CPU utilization of the \texttt{pytest} subprocess and its child processes. We then compute the average memory (MB) and average CPU utilization (\%) over the run. Let $M_{\text{gen}}, C_{\text{gen}}$ and $M_{\text{ref}}, C_{\text{ref}}$ denote the corresponding averages for the generated code and the reference implementation. These measurements provide a comparative runtime signal under a fixed environment. If the resource suite fails to execute successfully (e.g., failed tests), we set $\mathrm{Ru}=0$. Otherwise, we compute reference-aligned ratios and average memory and CPU when both are available:
\begin{equation}
\mathrm{Ru} \;=\;
\begin{cases}
\frac{1}{2}\!\left(
\min\!\left(1,\frac{M_{\text{ref}}}{M_{\text{gen}}}\right)
+
\min\!\left(1,\frac{C_{\text{ref}}}{C_{\text{gen}}}\right)
\right), & \text{CPU available},\\[6pt]
\min\!\left(1,\frac{M_{\text{ref}}}{M_{\text{gen}}}\right), & \text{memory only}.
\end{cases}
\label{eq:resource}
\end{equation}

% \noindent To reflect the fact that non-functional quality attributes correspond to qualitatively different engineering risks (e.g., security vulnerabilities and efficiency regressions), we determine attribute weights using the Analytic Hierarchy Process (AHP). 
\noindent We aggregate the five normalized dimensions with AHP-derived weights. We use AHP to provide a transparent and reproducible aggregation rule for multi-dimensional quality, not to claim a universal weighting scheme for all software domains. To avoid over-reliance on a single aggregate score, we also report the full per-dimension breakdown.
We ground the relative importance ordering on Botchway et al., who apply AHP to expert questionnaires and report maintainability and security as the highest-priority quality attributes, followed by robustness and efficiency \citep{botchway2021evaluating}. Based on this ordering, we construct a pairwise comparison matrix using the standard Saaty scale \citep{franek2014judgment}. We compute the normalized principal eigenvector to obtain the weight vector and verify consistency using the AHP consistency ratio (CR). In our setting, $\mathrm{CR}=0.030$ ($<0.1$), indicating an acceptable level of consistency. 
The full AHP derivation (matrix, eigenvector solution, CI/RI/CR) is documented in \href{https://anonymous.4open.science/r/RAL-Bench}
{\textbf{\textcolor{oai-blue}{our anonymous repository}}} for reproducibility.
The non-functional score aggregates five normalized quality dimensions using a weighted sum. Here, $M$, $S$, $Rb$, $E$, and $Ru$ denote the normalized scores of maintainability, security, robustness, efficiency, and resource usage, respectively:

\begin{equation}
\mathrm{NF}
= 0.36\,\mathrm{M}
+ 0.24\,\mathrm{S}
+ 0.16\,\mathrm{Rb}
+ 0.12\,\mathrm{E}
+ 0.12\,\mathrm{Ru}.
\label{eq:nf}
\end{equation}

% This non-functional score provides a more reliable and discriminative measure of the practical quality of generated applications, beyond functional correctness. 
This non-functional score provides a compact summary of multi-dimensional software quality and is reported together with the five per-dimension scores.
In Section 3.2, we show that it is both discriminative across LLMs and stable under reruns.

\textbf{Metric limitations.} These metrics are operational proxies designed for scalable benchmark evaluation. They do not replace expert audit or domain-specific quality assessment. In particular, static-analysis-based measures may produce false positives or false negatives, and runtime measures are environment-dependent. We therefore use them as comparative signals under a fixed protocol.

\subsection{Evaluation Pipeline}

\begin{figure}[t]
  \centering
  \includegraphics[width=0.75\columnwidth]{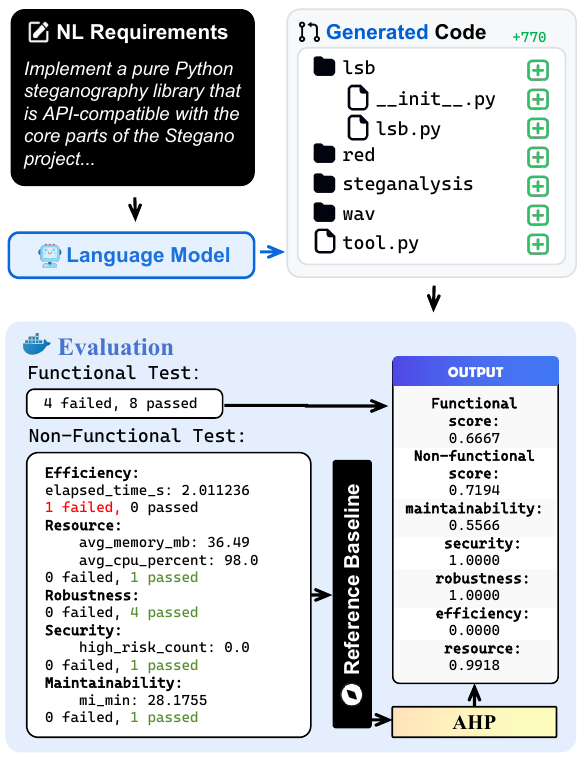}
  \caption{Evaluation pipeline overview. % \hy{re-organzine the layout, reduce empty space}
  }
  \vspace{-10px}
  \label{fig:eval-pipeline}
\end{figure}

We evaluate application-level code generation under a controlled end-to-end pipeline. The purpose of this setup is standardized comparison across models under a fixed generation and testing budget. It is not intended to represent all practical development workflows.
The evaluation workflow is illustrated in Fig. \ref{fig:eval-pipeline}.

\textbf{Input.} First, for each target project, we provide the LLM with a concise natural-language requirement that captures the core functionality to be implemented. To ensure strict interface alignment and avoid evaluation artifacts caused by mismatched entry points, we also provide the reference repository’s expected module and package surface, guiding the model to generate a repository that conforms to the same API boundary. 
% Given this requirement and the interface constraints, the LLM generates a complete repository from scratch.
Given the requirement and the interface constraints, the LLM generates a complete repository in a single controlled pass. We use this setup to isolate base repository-generation ability without adding confounding factors from agent design, iterative repair policy, or tool-use budget.

\textbf{Functional Correctness via System Tests.} We then evaluate functional correctness using system test suites. Specifically, we execute functional tests against the generated repository and compute the functional score as the pass ratio across all functional test cases, which directly measures whether the generated application behavior matches the reference implementation’s intended functionality. 

\textbf{The Role of Interface Constraints.} On \textsc{Stegano} with GPT-5.2, omitting the interface specification yields a functional score of 0.1667. Providing the interface specification increases the score to 0.6667. Inspecting the corresponding failure cases shows that this improvement primarily stems from alleviating test-interface mismatches (e.g., incorrect entry points or module paths) rather than from failing to satisfy the task requirements. 
% As a result, the evaluation more accurately reflects the LLM’s ability to produce an executable repository that satisfies the end-to-end functional requirements.
This example shows why interface constraints are included in the protocol. They reduce mismatches in entry points and module paths, so that the benchmark better measures application behavior rather than avoidable interface mismatches.

\textbf{Non-functional Measurement.} 
% Beyond functional correctness, we further assess non-functional quality attributes by running a set of tests and analyses that produce measurable signals used by our scoring functions. These include execution time for efficiency (e.g., $elapsed\_time\_s$), resource utilization statistics (e.g., $avg\_memory\_mb$ and $avg\_cpu\_percent$), static security findings (e.g., the number of high-risk issues $high\_risk\_count$), and maintainability indicators (e.g., the minimum maintainability index $mi\_min$).
% For robustness, we define it as the system’s ability to remain stable under stress and edge-case execution paths. We quantify robustness using the pass ratio on our robustness-oriented test suite.
% A higher score indicates the generated project remains stable under stress and edge-case execution paths.
Beyond functional correctness, we run the non-functional measurement pipeline defined in Section~2.3. This pipeline collects runtime and static-analysis signals for the five quality dimensions and converts them into comparable per-dimension scores.

% For robustness, we operationalize it as the ability to consistently pass robustness-oriented test cases, and quantify robustness by the corresponding robustness test pass ratio on the robustness suite, reflecting whether the generated system remains stable under stress and edge-case execution paths.

% \textbf{Robustness Scoring.} For robustness, we define it as the system’s ability to remain stable under stress and edge-case execution paths.We quantify robustness using the pass ratio on our robustness-oriented test suite.
% A higher score indicates the generated project remains stable under stress and edge-case execution paths.

\textbf{Reference-normalized Scoring.} Because different projects have very different scales, we compute a per-project baseline by running the non-functional suite on the reference repository. We then score each generated repository by comparing its non-functional metrics against this baseline, making results comparable across projects. Concretely, for efficiency, resource usage, security, and maintainability, we first run the same measurement pipeline on the reference repository to obtain a task-specific baseline for each dimension. 
We then evaluate the generated repository under the same environment and normalize each raw measurement against its baseline using fixed, dimension-specific rules, yielding a score in [0, 1]. 
% A higher score indicates closer-to-reference (or better-than-reference) behavior on that dimension. 
The reference repository serves as a task-specific anchor for comparison, not as a claim of global optimality. We cap normalized scores at 1 so that reaching or exceeding the reference level is rewarded, while preventing one unusually favorable raw measurement from dominating the aggregate score.
The exact scoring rules and normalization functions are defined in Section 2.3.
% Section 3.1

\textbf{Aggregation and Outputs.} Finally, we aggregate the five non-functional dimension scores into an overall non-functional index using an AHP-derived weight vector \citep{botchway2021evaluating}. 
% This produces a weighted assessment of non-functional quality attributes that emphasizes practically important attributes while remaining consistent across projects. The pipeline outputs both the functional score and the non-functional score, together with the per-dimension breakdown (maintainability, security, robustness, efficiency, and resource usage), enabling the subsequent analyses to localize bottlenecks.
We report both the aggregate non-functional score and the five per-dimension scores. This design supports comparison at the benchmark level while preserving diagnostic detail for later analysis.

\textbf{Protocol Boundary.} Although many real development workflows are iterative or agent-based, we keep the main benchmark setting single-turn and zero-shot to maintain comparability across models. Studying multi-turn or agentic generation under the same benchmark is an important next step, but requires additional protocol design.

\subsection{Benchmark Validation}

\begin{table}[t]
\centering
\caption{Summary statistics of executable test suites.}
\label{tab:benchmark_validation_stats}
\small
\setlength{\tabcolsep}{8pt}
\renewcommand{\arraystretch}{1.08}
\begin{tabular}{l c}
\toprule
\textbf{Metric} & \textbf{Value} \\
\midrule
\# Tasks & 38 \\
Functional tests (mean / median / min / max) & 11.03 / 11.5 / 10 / 19 \\
Robustness tests (mean) & 4.03 \\
Efficiency tests (mean) & 1.21 \\
Resource tests (mean) & 1.39 \\
Executable suite count (mean / median) & 17.66 / 18.0 \\
\bottomrule
\end{tabular}
\vspace{-10px}
\end{table}

\textbf{Test statistics and behavioral coverage.}
To further validate the benchmark construction, we summarize the executable test suites for all 38 tasks.
As shown in Table~\ref{tab:benchmark_validation_stats}, the benchmark does not rely on only a few checks per task.
% Across tasks, functional tests have mean/median/min/max of 11.03/11.5/10/19.
When robustness, efficiency, and resource suites are included, the executable suite count reaches mean/median of 17.66/18.0 per task.
We also inspect representative tasks from different scenarios and find that their functional suites cover multiple behavior categories rather than a single shallow invocation path.
These results provide additional evidence that \bench uses task-specific black-box suites with multi-category behavioral coverage.

\textbf{Defect-detection sanity check.}
Beyond reference-repository validation, we perform a defect-detection sanity check to assess whether the black-box suites can detect representative application-level failures.
We evaluate all tasks in \bench and inject two defects per task, yielding 76 injected defects in total.
As shown in Table~\ref{tab:defect_sanity_check}, the functional suites detect all injected defects, while the robustness suites additionally detect a subset of runtime-stability defects.
The union detection rate is 1.00.
The results provide additional evidence that the suites have meaningful defect-detection power for representative application-level failures.

\begin{table}[t]
\centering
\caption{Defect-detection sanity check on all tasks.}
\label{tab:defect_sanity_check}
\small
\setlength{\tabcolsep}{8pt}
\renewcommand{\arraystretch}{1.08}
\begin{tabular}{l c}
\toprule
\textbf{Metric} & \textbf{Value} \\
\midrule
\# Tasks & 38 \\
\# Injected defects & 76 \\
Functional detection rate & 1.00 \\
Robustness detection rate & 0.375 \\
Union detection rate & 1.00 \\
\bottomrule
\end{tabular}
\vspace{-10px}
\end{table}

%% file: Evaluation.tex
\section{Benchmarking on \bench}

We evaluate the performance of state-of-the-art (SOTA) LLMs on application-level code generation with \bench. Specifically, we aim to address the following research questions (RQs):

\begin{itemize}
    \item RQ1: What is the performance of current LLMs in application-level code generation?
    \item RQ2: Where do LLMs fail in application-level code generation?
    \item RQ3: What is the cost of application-level code generation with current LLMs?
    % \item RQ4: What strategies can improve application-level code generation? 
    \item RQ4: Are mainstream code generation strategies effective for application-level code generation?
\end{itemize}

% RQ1: What is the performance of current LLMs in application-level code generation?

% RQ2: Where do LLMs fail in application-level code generation?

% RQ3: What is the cost of application-level code generation with current LLMs?

% RQ4: Are mainstream code generation strategies effective for application-level code generation?

\subsection{Evaluation Setup}

\textbf{Benchmark Models.} We comprehensively evaluate 16 LLMs across diverse model families. For standard LLMs, we consider Gemini-2.5-Pro \citep{comanici2025gemini}, Gemini-3-Pro-Preview, GPT-3.5-Turbo, GPT-4o-2024-11-20 \citep{hurst2024gpt}, GPT-4-Turbo \citep{achiam2023gpt}, GPT-5 \citep{leon2025gpt}, GPT-5.2, Claude-Haiku-4.5, Claude-4.5-Sonnet, DeepSeek-V3-0324 \citep{liu2024deepseek}, and DeepSeek-V3.2 \cite{liu2025deepseek}. For thinking LLMs, we evaluate Gemini-2.5-Pro-Thinking, GPT-o1 \citep{jaech2024openai}, GPT-o3-2025-04-16, Claude-3.7-Sonnet-Thinking, and DeepSeek-R1 \citep{guo2025deepseek}. All evaluations are conducted in a zero-shot setting using each model’s default prompt template. 
For each task, LLMs output the entire repository in one pass, within the LLM’s context-window limits.
To ensure reproducibility, we repeat each evaluation three times under identical settings and report the mean score across runs for all metrics.

\input{RQ1}

\input{RQ2}

\input{RQ3}

\input{RQ4}

%% file: RQ1.tex
\subsection{RQ1: Capability Assessment}

\begin{table*}[t]
  \centering
  \fontsize{7.5}{11.5}\selectfont
  \setlength{\tabcolsep}{5pt}
  \caption{Evaluation scores reported as percentages (higher is better). Best values per column are in bold.  % \hy{the row space could be reduced a bit}
  }
  \renewcommand{\arraystretch}{1.05}

  % --- Group colors ---
  \definecolor{gemA}{RGB}{238,244,255}
  \definecolor{gemB}{RGB}{227,238,255}
  \definecolor{gptA}{RGB}{236,255,242}
  \definecolor{gptB}{RGB}{224,250,233}
  \definecolor{claA}{RGB}{255,242,236}
  \definecolor{claB}{RGB}{255,232,222}
  \definecolor{dsA}{RGB}{245,245,250}
  \definecolor{dsB}{RGB}{235,235,245}
  \definecolor{avgRow}{RGB}{207,254,246}

  % unified thin separator (thickness controlled here)
  \newcommand{\groupsep}{\specialrule{0.4pt}{2pt}{2pt}}

  \begin{tabularx}{0.8\textwidth}{
    >{\centering\arraybackslash}X
    >{\centering\arraybackslash}p{1.5cm}
    *{7}{S[table-format=2.2,table-column-width=0.85cm]}
  }
    \toprule
    \multirow{2}{*}{\textbf{LLM}} &
    \multirow{2}{*}{\textbf{Provider}} &
    \multicolumn{7}{c}{\textbf{Scores (\%)}} \\
    \cmidrule(lr){3-9}
    & & {\textbf{Func.}} & {\textbf{Non-func.}} & {\textbf{Maint.}} & {\textbf{Sec.}} &
        {\textbf{Robust.}} & {\textbf{Eff.}} & {\textbf{Res.}} \\
    \midrule

    % ===================== Gemini =====================
    \rowcolor{gemA}
    \textbf{Gemini-2.5-Pro}    & \ProvGoogle   & 27.63 & 35.15 & 26.27 & 54.39 & 41.39 & 24.84 & 25.28 \\
    \rowcolor{gemB}
    \textbf{Gemini-3-Pro-Preview}      & \ProvGoogle    & \textbf{43.86} & 50.57 & 31.50 & 76.75 & 66.36 & \textbf{43.69} & 41.23 \\
    \groupsep

    % ===================== GPT (version order) =====================
    \rowcolor{gptA}
    \textbf{GPT-3.5-Turbo}     & \ProvOpenAI    & 29.76 & 55.08 & 36.92 & \textbf{97.37} & 79.62 & 29.80 & 17.57 \\
    \rowcolor{gptB}
    \textbf{GPT-4o-2024-11-20}            & \ProvOpenAI   & 27.48 & 57.59 & \textbf{37.50} & 96.05 & \textbf{84.07} & 35.54 & 27.71 \\
    \rowcolor{gptA}
    \textbf{GPT-4-Turbo}       & \ProvOpenAI    & 36.42 & \textbf{58.13} & 34.90 & \textbf{97.37} & 78.32 & 42.24 & 38.31 \\
    \rowcolor{gptB}
    \textbf{GPT-5-2025-08-07}             & \ProvOpenAI   & 38.50 & 57.51 & 33.84 & 94.74 & 77.02 & 41.81 & 43.73 \\
    \rowcolor{gptA}
    \textbf{GPT-5.2}           & \ProvOpenAI   & 41.81 & 57.05 & 33.90 & 92.11 & 77.19 & 42.79 & \textbf{43.83} \\
    \groupsep

    % ===================== Claude =====================
    \rowcolor{claA}
    \textbf{Claude-Haiku-4.5-20251001}  & \ProvAnthropic & 34.74 & 56.40 & 34.52 & \textbf{97.37} & 79.66 & 37.19 & 28.32 \\
    \rowcolor{claB}
    \textbf{Claude-4.5-Sonnet} & \ProvAnthropic & 39.32 & 49.64 & 30.19 & 84.21 & 66.93 & 31.50 & 33.94 \\
    \groupsep

    % ===================== DeepSeek =====================
    \rowcolor{dsA}
    \textbf{DeepSeek-V3-0324}       & \ProvDeepSeek  & 24.37 & 46.34 & 32.29 & 81.58 & 65.89 & 22.90 & 15.36 \\
    \rowcolor{dsB}
    \textbf{DeepSeek-V3.2}     & \ProvDeepSeek  & 29.92 & 51.26 & 36.07 & 89.47 & 63.49 & 26.91 & 28.46 \\

    \midrule
  
    \textbf{Average} & -- &
    33.98 & 52.25 & 33.45 & 87.40 & 70.90 & 34.47 & 31.25 \\
    \bottomrule
  \end{tabularx}

  \label{tab:llm_scores}
\end{table*}

\begin{table*}[t]
  \centering
  \fontsize{7.5}{11.5}\selectfont
  \setlength{\tabcolsep}{5pt}
  \caption{Evaluation scores for thinking models reported as percentages (higher is better). Best values per column are in bold.}
  \renewcommand{\arraystretch}{1.05}

  % --- Group colors (same palette as your main table) ---
  \definecolor{gemA}{RGB}{238,244,255}
  \definecolor{gemB}{RGB}{227,238,255}
  \definecolor{gptA}{RGB}{236,255,242}
  \definecolor{gptB}{RGB}{224,250,233}
  \definecolor{claA}{RGB}{255,242,236}
  \definecolor{claB}{RGB}{255,232,222}
  \definecolor{dsA}{RGB}{245,245,250}
  \definecolor{dsB}{RGB}{235,235,245}
  \definecolor{avgRow}{RGB}{207,254,246}

  % unified thin separator
  \newcommand{\groupsep}{\specialrule{0.4pt}{2pt}{2pt}}

  \begin{tabularx}{0.8\textwidth}{
    >{\centering\arraybackslash}X
    >{\centering\arraybackslash}p{1.5cm}
    *{7}{S[table-format=2.2,table-column-width=0.83cm]}
  }
    \toprule
    \multirow{2}{*}{\textbf{LLM}} &
    \multirow{2}{*}{\textbf{Provider}} &
    \multicolumn{7}{c}{\textbf{Scores (\%)}} \\
    \cmidrule(lr){3-9}
    & & {\textbf{Func.}} & {\textbf{Non-func.}} & {\textbf{Maint.}} & {\textbf{Sec.}} &
        {\textbf{Robust.}} & {\textbf{Eff.}} & {\textbf{Res.}} \\
    \midrule

    % ===================== Gemini =====================
    \rowcolor{gemA}
    \textbf{Gemini-2.5-Pro-Thinking} & \ProvGoogle
      & \textbf{43.44} & \textbf{56.09} & 34.80 & 89.47 & \textbf{75.16} & \textbf{42.68} & \textbf{41.15} \\
    \groupsep

    % ===================== OpenAI (o1 -> o3) =====================
    \rowcolor{gptA}
    \textbf{GPT-o1} & \ProvOpenAI
      & 33.88 & 55.44 & \textbf{37.44} & 92.11 & 74.07 & 36.35 & 30.37 \\
    \rowcolor{gptB}
    \textbf{GPT-o3-2025-04-16} & \ProvOpenAI
      & 30.27 & 55.44 & 35.01 & \textbf{96.05} & 73.60 & 35.89 & 30.80 \\
    \groupsep

    % ===================== Claude =====================
    \rowcolor{claA}
    \textbf{Claude-3.7-Sonnet-Thinking} & \ProvAnthropic
      & 27.34 & 52.94 & 37.29 & 92.11 & 63.34 & 30.65 & 29.95 \\
    \groupsep

    % ===================== DeepSeek =====================
    \rowcolor{dsA}
    \textbf{DeepSeek-R1} & \ProvDeepSeek
      & 16.23 & 27.95 & 12.36 & 56.84 & 38.42 & 18.57 & 12.36 \\

    \midrule
 
    \textbf{Average} & -- &
    30.23 & 49.57 & 31.38 & 85.32 & 64.92 & 32.83 & 28.92 \\
    \bottomrule
  \end{tabularx}

  \label{tab:llm_scores_thinking}
\end{table*}

Table~\ref{tab:llm_scores} and Table~\ref{tab:llm_scores_thinking} summarize model performance on \bench. We make four observations.

\textbf{Functional correctness is the main bottleneck.} Across standard LLMs, the average functional score is 33.98\%. For thinking LLMs, the average is 30.23\%. No model exceeds a 45\% functional score. These results show that current LLMs still struggle to complete end-to-end application logic under our evaluation protocol.

\textbf{Non-functional scores are generally higher, but they do not offset functional failures.} The average non-functional score is 52.25\% for standard LLMs and 49.57\% for thinking LLMs. This suggests that some generated repositories behave reasonably once they can run, but many still fail to satisfy the core functional requirements. At the same time, non-functional scores provide useful additional discrimination. For example, Gemini-2.5-Pro and GPT-4o-2024-11-20 have similar functional scores (27.63\% vs.\ 27.48\%), but their non-functional scores differ substantially (35.15\% vs.\ 57.59\%). This difference is reflected in the per-dimension breakdown. These results show that non-functional evaluation adds information that functional pass rate alone does not capture.

\textbf{Efficiency and resource usage still vary across models.} Even when models achieve similar functional scores, their runtime behavior can differ noticeably. This suggests that producing a runnable repository is not the same as producing an efficient or resource-aware one.

\textbf{Thinking does not provide consistent gains.} Gemini-2.5-Pro-Thinking improves substantially over Gemini-2.5-Pro, including a functional score increase from 27.63\% to 43.44\%. However, this pattern is not consistent across providers. On average, thinking LLMs remain slightly below standard LLMs on both functional and non-functional scores. This indicates that additional deliberation alone is not yet a reliable way to improve application-level code generation.

\begin{table*}[t]
\centering
\caption{Rerun stability (\(K=5\) full reruns). For each LLM, we compute the standard deviation (Std) and coefficient of variation (CV) over \(K\) reruns per project, then summarize across 38 projects using median (med) and 95th percentile (p95). Runtime metrics are reported using CV.}
\label{tab:rerun-stability}
\setlength{\tabcolsep}{3pt}
\renewcommand{\arraystretch}{1.10}
\small

\resizebox{\textwidth}{!}{%
\begin{tabular}{lcc|
S[table-format=1.4]S[table-format=1.4]
S[table-format=1.4]S[table-format=1.4]|
S[table-format=1.4]S[table-format=1.4]
S[table-format=1.4]S[table-format=1.4]|
S[table-format=1.4]S[table-format=1.4]
S[table-format=1.4]S[table-format=1.4]
S[table-format=1.4]S[table-format=1.4]}
\toprule
\multirow{3}{*}{\textbf{Model}} & \multirow{3}{*}{\textbf{\(K\)}} & \multirow{3}{*}{\textbf{Proj.}} &
\multicolumn{4}{c|}{\textbf{Functional Score Stability}} &
\multicolumn{4}{c|}{\textbf{Non-functional Score Stability}} &
\multicolumn{6}{c}{\textbf{Runtime Metrics (CV)}} \\
\cmidrule(lr){4-7} \cmidrule(lr){8-11} \cmidrule(lr){12-17}
& & &
\multicolumn{2}{c}{\textbf{Std}} & \multicolumn{2}{c|}{\textbf{CV}} &
\multicolumn{2}{c}{\textbf{Std}} & \multicolumn{2}{c|}{\textbf{CV}} &
\multicolumn{2}{c}{\textbf{Time}} &
\multicolumn{2}{c}{\textbf{Memory}} &
\multicolumn{2}{c}{\textbf{CPU}} \\
\cmidrule(lr){4-5} \cmidrule(lr){6-7}
\cmidrule(lr){8-9} \cmidrule(lr){10-11}
\cmidrule(lr){12-13} \cmidrule(lr){14-15} \cmidrule(lr){16-17}
& & &
\textbf{med} & \textbf{p95} & \textbf{med} & \textbf{p95} &
\textbf{med} & \textbf{p95} & \textbf{med} & \textbf{p95} &
\textbf{med} & \textbf{p95} & \textbf{med} & \textbf{p95} & \textbf{med} & \textbf{p95} \\
\midrule
\textbf{Claude-4.5-Sonnet} & 5 & 38
& 0.0000 & 0.0000 & 0.0000 & 0.0000
& 0.0003 & 0.0009 & 0.0007 & 0.0026
& 0.0208 & 0.0501 & 0.0072 & 0.0134 & 0.0111 & 0.0270 \\
\textbf{Gemini-3-Pro-Preview} & 5 & 38
& 0.0000 & 0.0000 & 0.0000 & 0.0000
& 0.0003 & 0.0017 & 0.0010 & 0.0050
& 0.0319 & 0.0795 & 0.0080 & 0.0136 & 0.0137 & 0.0353 \\
\textbf{GPT-5.2} & 5 & 38
& 0.0000 & 0.0000 & 0.0000 & 0.0000
& 0.0002 & 0.0006 & 0.0005 & 0.0017
& 0.0198 & 0.0468 & 0.0074 & 0.0138 & 0.0098 & 0.0214 \\
\midrule
\textbf{Overall} & 5 & 38
& 0.0000 & 0.0000 & 0.0000 & 0.0000
& 0.0003 & 0.0012 & 0.0007 & 0.0028
& 0.0224 & 0.0557 & 0.0075 & 0.0138 & 0.0127 & 0.0270 \\
\bottomrule
\end{tabular}%
}
\vspace{-2mm}
\end{table*}

\textbf{Rerun Stability.} We assess metric stability by rerunning the full benchmark pipeline five times for three representative models: GPT-5.2, Gemini-3-Pro-Preview, and Claude-4.5-Sonnet. For each model, we compute the standard deviation (Std) and coefficient of variation (CV) across reruns for each project, and then summarize the results over all 38 projects using the median (med) and 95th percentile (p95).
Table~\ref{tab:rerun-stability} shows that the benchmark is stable. First, functional scores are fully reproducible: Std and CV are both 0 for all evaluated projects and models. Second, the aggregated non-functional score is also highly stable, with an overall Std of 0.0003 (med) and 0.0012 (p95), and an overall CV of 0.0007 (med) and 0.0028 (p95). Third, runtime metrics show small variation across reruns, which is expected under normal execution noise such as OS scheduling and caching effects. Across projects, the overall CV is 0.0224 (med) and 0.0557 (p95) for runtime, 0.0075 (med) and 0.0138 (p95) for memory, and 0.0127 (med) and 0.0270 (p95) for CPU. Because each non-functional dimension is normalized against a fixed reference baseline before aggregation, these small fluctuations do not change the model-level conclusions.

\begin{tcolorbox}[
  enhanced,
  colback=grey,
  colframe=teal!60!black,
  boxrule=0.6pt,
  arc=10pt,
  left=4mm,right=4mm,
  top=2mm,bottom=2mm,
  drop shadow={black!40!white},
]
\textbf{\textit{Answer to RQ1:}}
\textit{Current LLMs remain limited in application-level code generation under our evaluation protocol, primarily because of weak functional correctness in end-to-end execution. Non-functional metrics offer additional discrimination across models, but do not alter this conclusion. Thinking models help in some cases, but their gains are not consistent across providers.}
\end{tcolorbox}

% \begin{tcolorbox}[
%   enhanced,
%   colback=grey,                % 背景色
%   colframe=teal!60!black,       % 边框颜色
%   boxrule=0.6pt,                % 边框线宽
%   arc=10pt,                     % 圆角大小
%   left=4mm,right=4mm,           % 内边距
%   top=2mm,bottom=2mm,
%   drop shadow={black!40!white}, % 右下角阴影
% ]
% \textbf{\textit{Answer to RQ1:} }
% \textit{Application-level generation is constrained by functional requirements and non-functional quality attributes under end-to-end execution, rather than by producing a superficially runnable repository. Moreover, thinking LLMs are not a reliable way to improve application-level code quality. They can yield large gains for specific model families (e.g., Gemini) but do not produce consistent improvements across providers. This suggests that additional deliberation does not reliably improve application-level code quality.}

% \end{tcolorbox}

%% file: RQ2.tex
\subsection{RQ2: Failure Analysis}

\begin{figure}[t]
  \centering
  \includegraphics[width=0.45\textwidth]{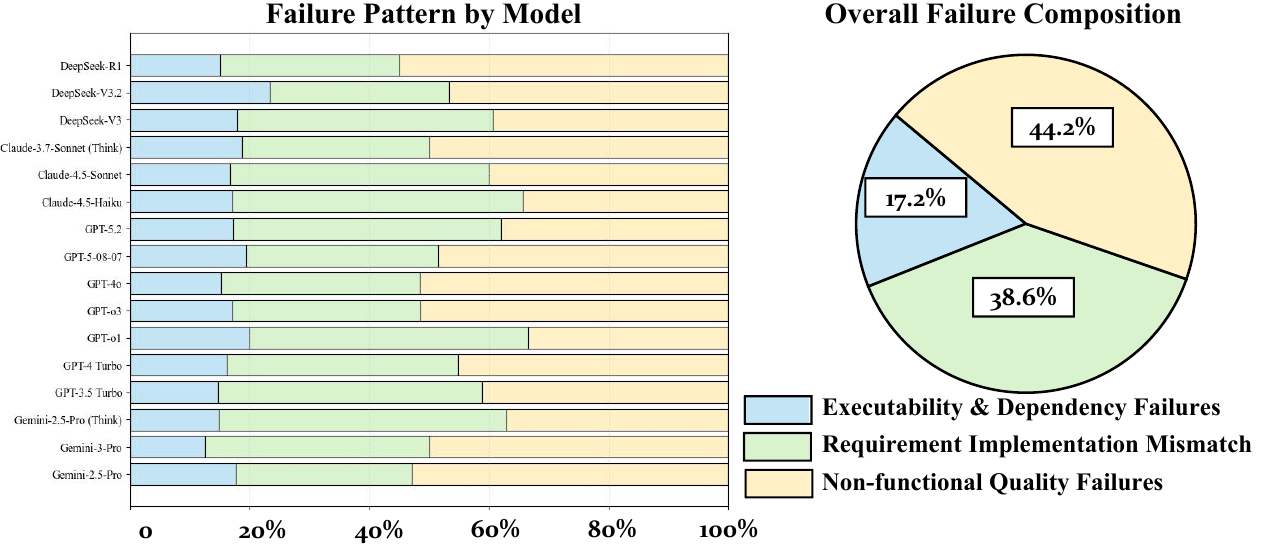}
  \caption{Failure-pattern distribution for application-level code generation.}
  \label{fig:rq2-failure-modes}
  \vspace{-10px}
\end{figure}

% \textbf{Setup.} To understand where and why current LLMs fail in application-level code generation, we conduct a fine-grained failure analysis based on our evaluation results. 
% We analyze 446 successfully generated repositories and over 4,500 test-case execution logs. Based on these artifacts, we construct a failure-pattern dataset for application-level code generation.
\textbf{Setup.} We analyze 446 generated repositories together with more than 4,500 test-case execution logs. For each failed case, we inspect the generated repository and its execution evidence, and assign one primary failure category. Based on this process, we construct a failure-pattern dataset for application-level code generation.

\textbf{Failure Taxonomy.} To characterize where LLMs fail in application-level code generation, we assign one primary failure category to each failed case. The taxonomy separates failures that prevent testing from those that arise during functional or runtime evaluation.

\noindent \textbf{1. \textcolor{linguistic}{Executability \& Dependency Failures.}} The generated repository cannot be built, imported, or collected for testing. Typical causes include missing dependencies, broken module paths, and malformed code structure. In this case, functional evaluation cannot start.

\noindent \textbf{2. \textcolor{egocentric}{Requirement--Implementation Mismatches.}} The tests execute, but the implementation does not satisfy the required application behavior. The repository is runnable, yet its outputs or behaviors do not match the tested requirements.

\noindent \textbf{3. \textcolor{relational}{Non-functional Quality Failures.}} Execution reaches the runtime stage, but the repository fails because of instability or inefficiency, such as exceptions, hangs, or timeouts. These failures reflect problems in robustness, liveness, or performance.

\textbf{Main Results.} Fig.~\ref{fig:rq2-failure-modes} shows the distribution of these three failure patterns. At the aggregate level, requirement--implementation mismatches and non-functional quality failures account for 82.8\% of all cases, while executability and dependency failures account for 17.2\%. This indicates that most failures occur after basic executability has been reached.

At the model level, Fig.~\ref{fig:rq2-failure-modes} (left) shows clear cross-model variation. Executability and dependency failures are consistently the smallest share, usually around 15\% to 20\%. By contrast, requirement--implementation mismatches and non-functional quality failures dominate for every model. However, models differ in which failure pattern is most common. Some are mainly limited by requirement--implementation mismatches after reaching executability, while others fail more often because of runtime instability or timeouts.

\begin{figure*}[t]
  \centering
  \includegraphics[width=0.75\textwidth]{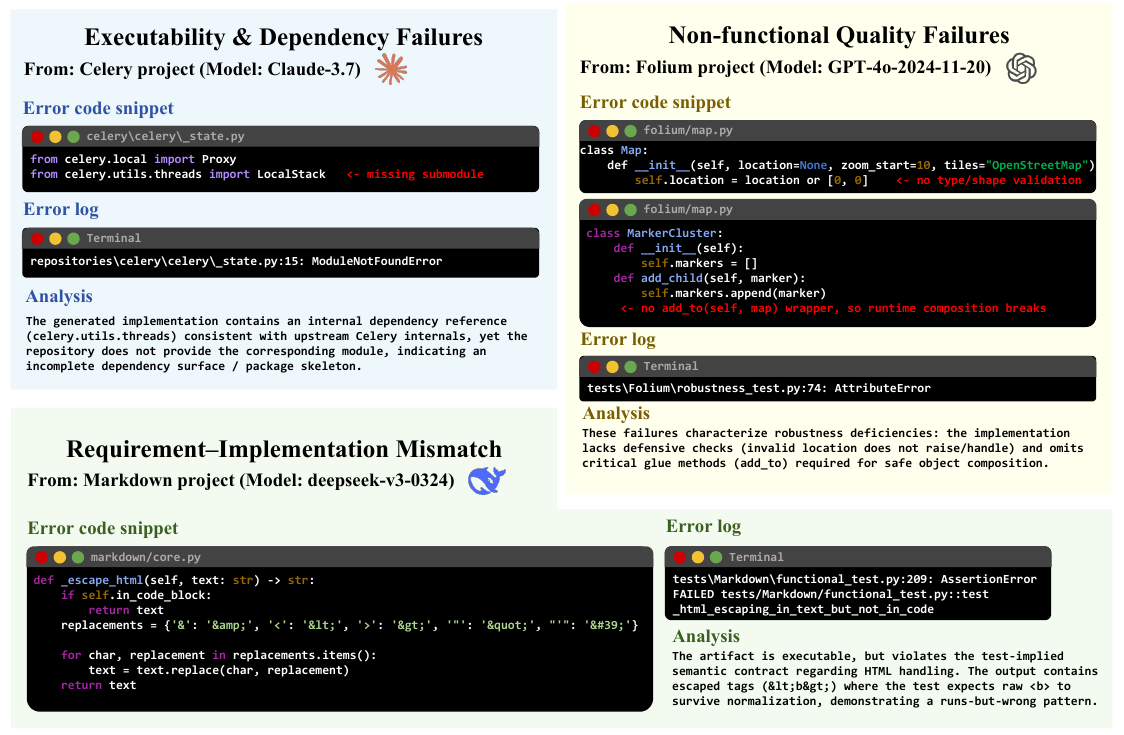}
  \caption{Case studies of three representative failure modes in application-level code generation.}
  \vspace{-7px}
  \label{fig:rq2-case-studies}
\end{figure*}

\textbf{Case Studies.} Fig.~\ref{fig:rq2-case-studies} presents three representative failures, one for each category. Together, these examples show that application-level generation can fail before testing starts, during functional verification, or during runtime robustness evaluation.

\textbf{In the executability and dependency failure case,} the Celery repository generated by Claude-3.7-Sonnet-Thinking fails during pytest collection. The code imports celery.utils.threads.LocalStack, but the referenced submodule is missing, which raises a ModuleNotFoundError. This is an executability failure. The repository does not expose a complete importable dependency surface, so functional verification cannot begin.

\textbf{In the requirement--implementation mismatch case,} the Markdown repository generated by DeepSeek-V3-0324 runs the tests but fails on assertions. Its HTML escaping routine produces escaped tags such as \texttt{\&lt;b\&gt;}, while the tests expect raw \texttt{<b>} to be preserved in specific contexts. The repository is executable, but its behavior does not match the tested requirement. As a result, it runs successfully but produces incorrect outputs.

\textbf{In the non-functional quality failure case,} the Folium repository generated by GPT-4o-2024-11-20 reaches runtime but breaks under robustness-oriented execution paths. Missing defensive checks and incomplete compositional interfaces, such as omitting \texttt{add\_to}, lead to runtime exceptions including \texttt{AttributeError}. This is a non-functional quality failure: the repository reaches runtime, but breaks under robustness-oriented execution paths.

\begin{tcolorbox}[
  enhanced,
  colback=grey,
  colframe=teal!60!black,
  boxrule=0.6pt,
  arc=10pt,
  left=4mm,right=4mm,
  top=2mm,bottom=2mm,
  drop shadow={black!40!white},
]
\textit{\textbf{Answer to RQ2: }Most failures in application-level code generation occur after basic executability has been reached. The main difficulty is no longer dependency resolution, but satisfying end-to-end functional requirements and maintaining stable runtime behavior. Models differ mainly in which of these two post-executability bottlenecks appears first.}
\end{tcolorbox}

%% file: RQ3.tex
\subsection{RQ3: Cost Analysis}

% To better understand the practical cost of application-level code generation, we perform a one-round generation for each task without iterative refinement. We compute costs using providers’ input and output token pricing and report the mean cost over three independent benchmark runs. 
To measure the generation cost of application-level code generation, we use a single-round generation setting without iterative refinement. We estimate monetary cost from provider-reported input and output token pricing, and report the mean generation cost over three independent benchmark runs. These numbers reflect model-generation cost only, not downstream test execution cost.
% Table \ref{tab:avg_cost} reports the average per-run monetary cost of evaluating each LLM on \bench. Our key observations are as follows:
Table~\ref{tab:avg_cost} reports the average per-run generation cost of each LLM on \bench. We make three observations.

\begin{table}[t]
  \centering
  \fontsize{7.5}{11.0}\selectfont
  \setlength{\tabcolsep}{5pt}
  \renewcommand{\arraystretch}{1.10}

  \caption{Average per-run generation cost (USD) for standard and thinking LLMs.}
  \label{tab:avg_cost}

  % --- colors ---
  \definecolor{gemA}{RGB}{238,244,255}
  \definecolor{gemB}{RGB}{227,238,255}
  \definecolor{gptA}{RGB}{236,255,242}
  \definecolor{gptB}{RGB}{224,250,233}
  \definecolor{claA}{RGB}{255,242,236}
  \definecolor{claB}{RGB}{255,232,222}
  \definecolor{dsA}{RGB}{245,245,250}
  \definecolor{dsB}{RGB}{235,235,245}
  \definecolor{sectionRow}{RGB}{240,240,240}

  % ===================== Standard =====================
  \begin{minipage}[t]{\linewidth}
    \centering
    \begin{tabularx}{0.8\linewidth}{
      >{\raggedright\arraybackslash}p{3cm}
      >{\raggedright\arraybackslash}X
      >{\raggedleft\arraybackslash}p{1cm}
    }
      \toprule
      \rowcolor{sectionRow}
      \multicolumn{3}{c}{\textbf{Standard LLMs}} \\
      \midrule
      \textbf{LLM} & \textbf{Provider} & \textbf{Avg.\ (\$)} \\
      \midrule
      \rowcolor{gemA} \textbf{Gemini-2.5-Pro}       & \ProvGoogle    & \textbf{12.11} \\
      \rowcolor{gemB} \textbf{Gemini-3-Pro-Preview} & \ProvGoogle    & 8.14 \\
      \rowcolor{gptA} \textbf{GPT-3.5-Turbo}        & \ProvOpenAI    & 0.54 \\
      \rowcolor{gptB} \textbf{GPT-4o-2024-11-20}    & \ProvOpenAI    & 0.42 \\
      \rowcolor{gptA} \textbf{GPT-4-Turbo}          & \ProvOpenAI    & 2.89 \\
      \rowcolor{gptB} \textbf{GPT-5-2025-08-07}                & \ProvOpenAI    & 4.04 \\
      \rowcolor{gptA} \textbf{GPT-5.2}              & \ProvOpenAI    & 2.67 \\
      \rowcolor{claA} \textbf{Claude-Haiku-4.5}     & \ProvAnthropic & 0.95 \\
      \rowcolor{claB} \textbf{Claude-4.5-Sonnet}    & \ProvAnthropic & 11.84 \\
      \rowcolor{dsA}  \textbf{DeepSeek-V3-0324}     & \ProvDeepSeek  & 1.34 \\
      \rowcolor{dsB}  \textbf{DeepSeek-V3.2}        & \ProvDeepSeek  & 0.15 \\
      \midrule
      \textbf{Average} & \textemdash & \textbf{4.10} \\
      \bottomrule
    \end{tabularx}
  \end{minipage}

  \vspace{5mm}

  % ===================== Thinking =====================
  \begin{minipage}[t]{\linewidth}
    \centering
    \begin{tabularx}{0.8\linewidth}{
      >{\raggedright\arraybackslash}p{3.4cm}
      >{\raggedright\arraybackslash}X
      >{\raggedleft\arraybackslash}p{0.9cm}
    }
      \toprule
      \rowcolor{sectionRow}
      \multicolumn{3}{c}{\textbf{Thinking LLMs}} \\
      \midrule
      \textbf{LLM} & \textbf{Provider} & \textbf{Avg.\ (\$)} \\
      \midrule
      \rowcolor{gemA} \textbf{Gemini-2.5-Pro-Thinking}    & \ProvGoogle    & 8.07 \\
      \rowcolor{gptA} \textbf{GPT-o1}                     & \ProvOpenAI    & 14.05 \\
      \rowcolor{gptB} \textbf{GPT-o3-2025-04-16}          & \ProvOpenAI    & 7.72 \\
      \rowcolor{claA} \textbf{Claude-3.7-Sonnet-Thinking} & \ProvAnthropic & \textbf{16.63} \\
      \rowcolor{dsA}  \textbf{DeepSeek-R1}                & \ProvDeepSeek  & 8.52 \\
      \midrule
      \textbf{Average} & \textemdash & \textbf{11.00} \\
      \bottomrule
    \end{tabularx}
  \end{minipage}

  \vspace{-1mm}
\end{table}

% \textbf{Cost variability.} Per-run costs range from $0.15 (DeepSeek-V3.2) to $16.63 (Claude-3.7-Sonnet-Thinking), a difference of more than 100$\times$. This indicates that the cost of application-level code generation can differ substantially depending on the chosen model. 
\textbf{Cost variability.} Per-run generation cost ranges from \$0.15 for DeepSeek-V3.2 to \$16.63 for Claude-3.7-Sonnet-Thinking, a difference of more than 100$\times$. This shows that application-level generation cost varies substantially across models.

% \textbf{Standard vs. thinking cost.} On average, thinking LLMs incur higher per-run costs (\$11.00) than standard LLMs (\$4.10). Nonetheless, exceptions exist (e.g., Gemini-2.5-Pro-Thinking vs. Gemini-2.5-Pro), suggesting that the observed cost reflects both provider-specific pricing and realized input/output token volumes. The results also show that thinking LLMs consume more tokens to solve the same application-level generation tasks. This directly translates into higher monetary cost. However, this higher cost does not yield a consistent improvement in functional correctness, as shown in Tables \ref{tab:llm_scores} and \ref{tab:llm_scores_thinking}. This highlights that enabling LLMs to reason more effectively for application-level code generation remains a key challenge.

\textbf{Standard vs.\ thinking cost.} On average, thinking LLMs are more expensive per run than standard LLMs (\$11.00 vs.\ \$4.10). However, this pattern is not uniform across providers. For example, Gemini-2.5-Pro-Thinking costs less than Gemini-2.5-Pro. This suggests that observed cost depends on both provider pricing and realized token usage.

\textbf{Cost--performance mismatch.} Higher generation cost does not translate into consistent gains in functional correctness. As shown in Tables~\ref{tab:llm_scores} and \ref{tab:llm_scores_thinking}, more expensive models do not reliably achieve higher application-level functional scores. This pattern is especially visible for thinking models, which often use more tokens but do not consistently outperform standard models.

\begin{tcolorbox}[
  enhanced,
  colback=grey,                % 背景色
  colframe=teal!60!black,       % 边框颜色
  boxrule=0.6pt,                % 边框线宽
  arc=10pt,                     % 圆角大小
  left=4mm,right=4mm,           % 内边距
  top=2mm,bottom=2mm,
  drop shadow={black!40!white}, % 右下角阴影
]
\textbf{\textit{Answer to RQ3:} }
\textit{
% Application-level code generation shows a clear mismatch between cost and performance. Across models, the monetary cost per run differs by more than 100$\times$, but paying more does not reliably lead to better application-level functional correctness. Thinking LLMs usually generate many more tokens, which increases cost substantially. Yet this extra computation does not consistently improve application-level requirement satisfaction.
Application-level code generation shows a clear cost--performance mismatch. Per-run generation cost varies by more than 100$\times$ across models, but higher cost does not reliably lead to better functional performance. Thinking models are usually more expensive, yet they do not provide consistent gains under our evaluation protocol.
}

\end{tcolorbox}

%% file: RQ4.tex
\begin{table}[t]
  \centering
  \fontsize{8}{11.0}\selectfont
  \caption{Impact of automated generation strategies on application-level quality (GPT-5.2). Scores are reported as percentages (\%).}
  \label{tab:rq4_strategies}
  \setlength{\tabcolsep}{5pt}
  \renewcommand{\arraystretch}{1.12}

  % --- row colors (each row different) ---
  \definecolor{rowA}{RGB}{236,255,242} % light green
  \definecolor{rowB}{RGB}{238,244,255} % light blue
  \definecolor{rowC}{RGB}{255,242,236} % light orange
  \definecolor{rowD}{RGB}{245,245,250} % light gray-blue
  \definecolor{rowE}{RGB}{255,232,222} % light peach

  \sisetup{detect-weight=true, detect-inline-weight=math}

  % IMPORTANT: use \linewidth and DO NOT strip outer padding with @{}
\begin{tabularx}{\linewidth}{
    >{\raggedright\arraybackslash}X
    >{\centering\arraybackslash}S[table-format=3]
    >{\centering\arraybackslash}S[table-format=3.1]
    >{\centering\arraybackslash}S[table-format=3.1]
    >{\centering\arraybackslash}S[table-format=3.1]
    >{\centering\arraybackslash}S[table-format=3.1]
    >{\centering\arraybackslash}S[table-format=3.1]
    >{\centering\arraybackslash}S[table-format=3.1]
  }
    \toprule
    \textbf{Strat.} &
    \textbf{Func.} &
    \textbf{Non-func.} &
    \textbf{Maint.} &
    \textbf{Sec.} &
    \textbf{Rob.} &
    \textbf{Eff.} &
    \textbf{Res.} \\
    \midrule

    % \rowcolor{rowA}

    \rowcolor{rowB}
    \textbf{S1 }     & 37.70 & 46.10 & 35.30 & 78.90 & 63.70 & 15.00 & 20.40 \\
\rowcolor{rowA}
\textbf{S2 }   & 40.20 & 49.90 & 36.10 & 84.20 & 67.60 & 21.00 & 28.00 \\
\rowcolor{rowE}
\textbf{S3 }     & 41.20 & 49.60 & 35.60 & 86.80 & 69.60 & 14.40 & 25.30 \\
\textbf{Baseline}                          & 41.80 & 57.10 & 33.90 & 92.10 & 77.20 & 42.80 & 43.80 \\
    \bottomrule

      % \vspace{-12px}
  \end{tabularx}

\vspace{-5mm}

\end{table}

% \subsection{RQ4: Are mainstream code generation strategies effective for application-level code generation?}

\begin{figure*}[t]
  \centering
  \includegraphics[width=0.77\linewidth]{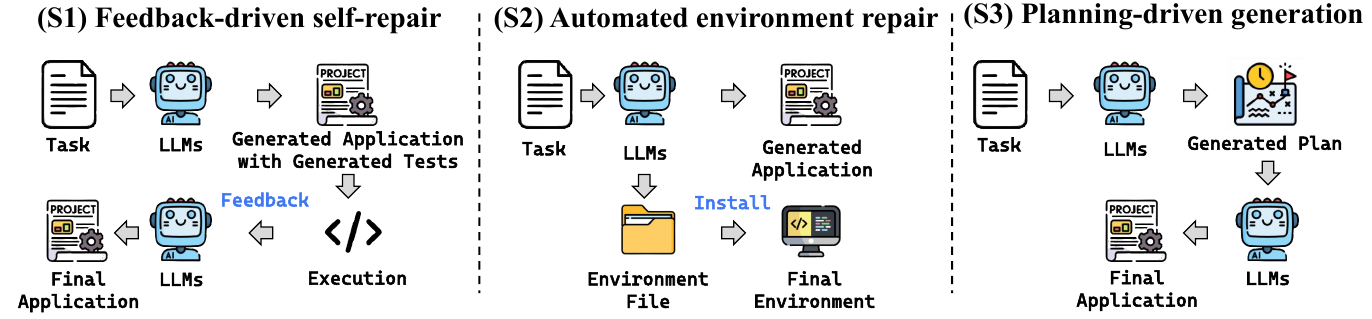}
  \caption{Overview of the three automated generation strategies evaluated in RQ4.}
  \vspace{-7px}
  \label{fig:rq4_strategies_overview}
\end{figure*}

\subsection{RQ4: Strategy Evaluation}

\textbf{Setup.} To examine whether common automated code generation strategies help in application-level generation, we conduct a controlled comparison on GPT-5.2 under the same evaluation pipeline. This experiment is intended as a single-model controlled study rather than a universal ranking of all strategy families. We use GPT-5.2 as the base model because it shows strong baseline performance at moderate cost, which makes repeated controlled comparison practical. We evaluate three representative one-shot strategies: \textbf{(S1)} feedback-driven self-repair, which performs one repair step using generated tests;\textbf{ (S2)} automated environment repair, which patches missing dependencies and build artifacts; and \textbf{(S3)} planning-driven generation, which adds an explicit plan before staged implementation. Fig.~\ref{fig:rq4_strategies_overview} summarizes their end-to-end workflows.

\textbf{Results.} Table~\ref{tab:rq4_strategies} shows that the evaluated one-shot strategies do not yield reliable gains over the baseline in our controlled setting. The baseline remains the strongest overall, with 41.80\% functional correctness and 57.10\% non-functional quality. The most consistent pattern is the drop in non-functional quality: all three strategies score substantially below the baseline on the aggregate non-functional metric.
\textbf{S1: Feedback-driven self-repair.} S1 reduces the functional score to 37.70\% and also lowers the non-functional score to 46.10\%. This suggests that one-step repair based on generated tests is not well aligned with application-level failures, where behaviors are distributed across modules and execution conditions. In this setting, the repair target can be incomplete or biased, which may introduce regressions.
\textbf{S2: Automated environment repair.} S2 reaches 40.20\% functional correctness and 49.90\% non-functional quality. This suggests that patching dependencies or build artifacts alone is not enough. As shown in RQ2, many remaining failures occur after executability has already been reached. Environment repair does not address missing end-to-end behaviors or unstable runtime interactions.
\textbf{S3: Planning-driven generation.} S3 does not improve functional correctness over the baseline (41.20\% vs.\ 41.80\%) and substantially reduces non-functional quality (49.60\% vs.\ 57.10\%). This suggests that an upfront one-shot plan is not enough to capture the distributed constraints of application-level tasks, such as cross-module interactions, configuration, and runtime behavior.

\begin{tcolorbox}[
  enhanced,
  colback=grey,
  colframe=teal!60!black,
  boxrule=0.6pt,
  arc=10pt,
  left=4mm,right=4mm,
  top=2mm,bottom=2mm,
  drop shadow={black!40!white},
]
\textbf{\textit{Answer to RQ4:}}
\textit{In our controlled GPT-5.2 study, the evaluated one-shot strategies do not provide reliable gains over the baseline. Their main limitation is not only weak improvement in functional correctness, but also a clear drop in non-functional quality. This suggests that isolated one-step interventions are not sufficient for application-level generation.}
\end{tcolorbox}

% \begin{table*}[t]
% \centering
% \caption{Practical workflow comparison on the 35-task overlap.}
% \label{tab:robust_practical_workflows}
% \small
% \setlength{\tabcolsep}{7pt}
% \renewcommand{\arraystretch}{1.08}
% \begin{tabular}{l c c c c c c c}
% \toprule
% \textbf{Setting} & \textbf{Tasks} & \textbf{Mean Func.} & \textbf{Mean NF} & $\Delta$ \textbf{Func.} & $\Delta$ \textbf{NF} & \textbf{Improved} & \textbf{Worse} \\
% \midrule
% Baseline              & 35 & 0.4539 & 0.6195 & --      & --      & -- & -- \\
% s1                    & 35 & 0.4096 & 0.5005 & -0.0443 & -0.1190 & 9  & 9  \\
% s2                    & 35 & 0.4366 & 0.5420 & -0.0174 & -0.0774 & 6  & 9  \\
% s3                    & 35 & 0.4468 & 0.5380 & -0.0071 & -0.0815 & 6  & 10 \\
% Best-of-strategies    & 35 & 0.5732 & 0.6533 & +0.1193 & +0.0338 & 12 & 0  \\
% \bottomrule
% \end{tabular}
% \end{table*}

\begin{table}[t]
\centering
\caption{Robustness of non-functional aggregation under alternative settings.}
\label{tab:robust_metric}
\small
\setlength{\tabcolsep}{7pt}
\renewcommand{\arraystretch}{1.08}
\begin{tabular}{l c c}
\toprule
\textbf{Variant} & \textbf{Spearman $\rho$} & \textbf{Top-3 Overlap} \\
\midrule
Equal weights                  & 0.9765 & 3 \\
No-cap perf./resource          & 0.9882 & 3 \\
Leave out maintainability      & 0.9294 & 2 \\
Leave out security             & 0.9941 & 3 \\
Leave out robustness           & 0.9500 & 3 \\
Leave out performance          & 0.9676 & 3 \\
Leave out resource             & 0.9029 & 3 \\
Rank-based aggregate           & 0.6441 & 1 \\
\bottomrule
\end{tabular}
\vspace{-10px}
\end{table}

\subsection{Robustness Analyses}

\textbf{Prompt sensitivity.}
We conduct a targeted prompt-sensitivity study on GPT-5.2 over 38 tasks, using three requirement variants with three repeats each.
Prompt wording affects application-level performance, but the aggregate differences are small.
The mean functional scores are 0.4181 for the baseline, 0.4972 for the clarified variant, and 0.4919 for the paraphrased variant.
These results suggest that prompt variation does not change the main finding that functional correctness remains the dominant bottleneck.

% \textbf{Practical workflow comparison.}
% We compare the zero-shot baseline with simple practical variants derived from our existing strategy pool.
% As shown in Table~\ref{tab:robust_practical_workflows}, no single strategy consistently outperforms the baseline across the 35 overlapping tasks.
% However, an optimistic best-of-strategies upper bound improves the mean functional score from 0.4539 to 0.5732 and the mean non-functional score from 0.6195 to 0.6533.
% These results suggest that the zero-shot benchmark remains a useful controlled baseline, while also leaving room for more adaptive multi-step workflows.

% \textbf{Public-repository exposure sanity check.}
% Because \bench uses public GitHub repositories, we cannot rule out pretraining exposure.
% We therefore perform a simple popularity and metadata sanity check.
% Repository popularity does not show a strong positive association with benchmark performance: for example, the correlation between strong-model functional score and $\log_{10}(\text{stars}+1)$ is weak and negative (Pearson $r=-0.3082$, $p=0.0634$), and higher-star repositories show lower mean functional performance than lower-star repositories (0.3965 vs.\ 0.5518).
% These results do not eliminate the possibility of exposure, but they do not suggest that performance is dominated by repository popularity alone.

\textbf{Metric robustness.}
We test the robustness of the non-functional aggregate under alternative weighting and normalization choices.
As shown in Table~\ref{tab:robust_metric}, the model ranking remains highly stable for reasonable variants, including equal weights, uncapped performance/resource normalization, and leave-one-dimension-out settings.
Only a substantially different rank-based aggregate leads to a noticeably lower rank correlation.
Overall, these results suggest that the main findings are not driven by a fragile choice of non-functional aggregation.

%% file: Discussion.tex
\section{Discussion}

\textbf{Application-level generation is different from snippet-level generation.} Our results show that strong snippet-level generation does not directly transfer to application-level generation. In \bench, the main difficulty is not producing locally plausible code, but producing a repository that works end to end. This requires consistent interfaces across modules, correct dependency setup, stable runtime behavior, and satisfaction of application-level functional requirements.

\textbf{One-shot interventions have limited effect in our controlled setting.} Our controlled GPT-5.2 study in Section~3.5 shows that the evaluated one-shot strategies do not provide reliable gains over the baseline. In particular, they often reduce non-functional quality even when functional scores change little. This result is consistent with our failure analysis: many failures occur after executability has been reached, at the stage of requirement satisfaction or runtime robustness. A single repair, environment patch, or upfront plan is therefore often not enough.

\textbf{Possible next steps.} These findings suggest that future work may benefit from workflows that make requirements more explicit, coordinate changes across modules, and monitor non-functional quality during generation. We highlight three possible directions. 
\circnum{1}\ \textbf{Application-level requirement analysis.} Future systems may benefit from turning a short natural-language requirements into a clearer task specification before generation. This specification can include expected behaviors, outputs, and important edge cases. A more explicit requirement representation may help reduce later requirement--implementation mismatches.
\circnum{2}\ \textbf{Cross-module repair.} Many application-level failures involve more than one file or module. Future work can therefore study repair methods that use application-level feedback to update related files together, rather than patching one local issue at a time. This may be especially useful when the failure comes from interface mismatch, missing coordination across modules, or unstable runtime interactions.
\circnum{3}\ \textbf{Quality-aware generation.} Future systems may also treat robustness, efficiency, resource usage, maintainability, and security as explicit constraints during generation. This may help detect quality regressions earlier, especially at module boundaries and under runtime execution. Such a direction is relevant because non-functional quality remains weak even when basic executability has been reached.

%% file: Related_Work.tex
\section{Related Work}

\textbf{LLMs for code.} Code generation has been long studied with recent approaches focused on pretrained LLMs for code, prompt engineering, retrieval-augmented generation, and agents \citep{jiang2025agentic, jiang2024survey}. Firstly, powerful foundation models for code (e.g., CodeX \citep{chen2022codet}, Code Llama \citep{roziere2023code}, StarCoder \citep{li2023starcoder}) are built by pretraining on large-scale code and natural language corpora, sometimes followed by instruction or domain-specific fine-tuning. On top of these models, prompt engineering techniques such as CodeChain \citep{le2023codechain}, SCoT \citep{li2025structured} and MoT \citep{pan2025modularization} aim to improve functional correctness without changing model parameters. To support repository-level understanding, retrieval-augmented methods like REPOFUSE \citep{liang2024repofuse}, DraCo \citep{cheng2024dataflow} and GraphCode \citep{russold2024graphcode} retrieve relevant files or snippets and feed them as additional context so that generated code better aligns with existing projects. More recently, agentic frameworks embed LLMs into interactive development environments where models can iteratively plan, edit code, run tests and incorporate execution feedback, with multi-agent variants assigning specialized roles to further improve robustness and code quality \citep{liu2024large, yao2022react, shinn2023reflexion}.

\textbf{Coding benchmarks for LLMs.}
Benchmarks for code generation are the primary tools for measuring and comparing the coding abilities of LLMs. For functional requirements, researchers have proposed function-level (e.g., HumanEval \citep{chen2021evaluating}, MBPP \citep{austin2021program}, EvalPlus \citep{liu2023your}), contest-level (e.g., APPS \citep{hendrycks2021measuring}, LiveCodeBench \citep{jain2024livecodebench}), class-level (e.g., ClassEval \citep{du2023classeval}), repository-level (e.g., RepoEval \citep{lozhkov2024starcoder}, CoderEval \citep{yu2024codereval}, DevEval \citep{li2024deveval}), and feature-level (e.g., FEA-Bench \citep{li2025fea}, NoCodeBench \citep{deng2025nocode}, FeatBench \citep{chen2025featbench}) benchmarks, all of which rely on execution-based metrics such as pass@k.
Beyond functional correctness, several benchmarks, such as EvalPerf \citep{liu2024evaluating} and EffiBench \citep{huang2024effibench}, begin to assess the efficiency of LLM-generated code by additionally measuring runtime and resource usage. However, these benchmarks still operate mostly at the function or contest level, or focus on modifying existing repositories, and thus do not evaluate end-to-end application generation. In contrast, our work targets application-level code generation, where LLMs must synthesize complete multi-file projects from scratch given only natural-language requirements. We further evaluate the generated programs using black-box system tests that simultaneously capture functional correctness and non-functional properties against real-world reference repositories.

%% file: Conclusion_and_Future_Work.tex
\section{Conclusion and Future Work}

% We present \bench - ...

% We study how models generate application-level code by building \bench and characterizing the performance and behavior of code LLMs on it. \bench grounds each task in a widely used GitHub project. We distill a concise natural-language requirement, construct black-box system tests that cover functional correctness and ISO/IEC 25010-inspired non-functional attributes, and validate each test case on the reference repository to ensure oracle soundness and end-to-end executability. This enables rigorous measurement of whether an LLM can synthesize a runnable repository that satisfies system-level functional requirements while maintaining practically relevant software quality.

We introduce \bench, an application-level benchmark grounded in widely used GitHub projects that evaluates end-to-end functional correctness and ISO/IEC 25010-inspired non-functional quality attributes with reference-validated oracles. 
Our evaluation across 16 frontier LLMs reveals a consistent capability gap between current LLMs and real-world application-level code generation.
% , which LLMs can generate a runnable project with proper structure, dependency management.
Functional correctness remains the dominant bottleneck. No LLM exceeds a 45\% functional score. 
By analyzing 446 generated repositories and over 4,500 execution logs, we further quantify that requirement–implementation mismatch and non-functional quality failures account for 82.8\% of failures, while executability and dependency failures contribute 17.2\%. 
We find that higher-cost reasoning variants and common code generation strategies do not reliably improve application-level generation and often degrade non-functional quality attributes. 
Going forward, we will expand \bench to broader projects and settings. Also, we will develop more effective workflows to better align generation with application-level functional requirements and non-functional quality attributes.

\section{Data Availability}

To ensure the reproducibility of our results and to provide transparency in our research, we have made all related scripts and data publicly available. All resources are available at 
% \href{https://anonymous.4open.science/r/RAL-Bench}
%      {\textbf{\textcolor{oai-orange}{\nolinkurl{https://anonymous.4open.science/r/RAL-Bench}}}} and are also archived in \href{https://zenodo.org/records/19243349}
% {\textbf{\textcolor{oai-orange}{an anonymous Zenodo repository}}}.
\href{https://github.com/Wwstarry/RAL-Bench}
     {\textbf{\textcolor{oai-orange}{\nolinkurl{https://github.com/Wwstarry/RAL-Bench}}}}.